\documentclass[11pt,british,english]{article}
\pdfoutput=1
\usepackage[T1]{fontenc}
\usepackage[latin9]{inputenc}
\usepackage{geometry}
\usepackage{color}
\usepackage{amsmath}
\usepackage{amssymb}
\usepackage{subfig}
\usepackage{esint}
\usepackage{cite}
\usepackage{float}
\usepackage[colorlinks=true,linkcolor=blue, citecolor=red, urlcolor=blue, bookmarks]{hyperref}

\makeatletter

\providecommand{\tabularnewline}{\\}

\numberwithin{equation}{section}
\numberwithin{figure}{section}
\numberwithin{table}{section}

\makeatother

\usepackage{babel}
\begin{document}

\title{\bf Symmetry resolved entanglement in integrable field theories via form factor bootstrap}

\author{Dávid X. Horváth$^{1}$ and Pasquale Calabrese$^{1,2}$~\\
 $^{1}${\small{}SISSA and INFN Sezione di Trieste, via Bonomea 265,
34136 Trieste, Italy}.\\
 $^{2}${\small{}International Centre for Theoretical Physics (ICTP),
Strada Costiera 11, 34151 Trieste, Italy.} }
\maketitle
\begin{abstract}
We consider the form factor bootstrap approach of integrable field theories to derive matrix elements of composite branch-point twist
fields associated with symmetry resolved entanglement entropies. The bootstrap equations are determined in an intuitive way and their solution is
presented for the massive Ising field theory and for the genuinely interacting sinh-Gordon model, both possessing a $\mathbb{Z}_{2}$ symmetry. 
The solutions are carefully cross-checked by performing various limits and by the application of the $\Delta$-theorem. 
The issue of symmetry resolution for discrete symmetries is also discussed.
We show that entanglement equipartition is generically expected and we identify the first subleading term (in the UV cutoff) breaking it.
We also present the complete computation of the symmetry resolved von Neumann entropy  
for an interval in the ground state of the paramagnetic phase of the Ising model. 
%Our findings are cross-checked against known results for the Ising spin chain. 
In particular, we compute the universal functions entering in the charged and symmetry resolved entanglement.
\end{abstract}

\baselineskip 18pt
\thispagestyle{empty}
\newpage

%\begin{center}
%\textbf{\LARGE {Trace distance in conformal field theory and spin chain}}
%\end{center}

\tableofcontents

\section{Introduction}

Symmetries play a central role in physics and in our understanding of nature. 
They are important guiding principle when formulating theories, their presence or absence or their breaking
have profound consequences on the physical properties of models and real-world systems; 
last but not least symmetries often provide a larger view in the description of the systems of interest. 
From a practical perspective, the presence of a symmetry usually leads to some kind of simplifications. 
In particular, for a quantum system the operator corresponding to the symmetry
commutes with the Hamiltonian and hence the two operators have common
eigenvectors or, in other words, the eigenstates of the system can
be characterised by quantum numbers associated with the symmetry operation.
The idea of exploiting the additional structures imposed by symmetry
for various physical objects is very fruitful and has been  recently
extended to the study of entanglement too.

When a system is in a pure state, the bipartite entanglement of a subsystem A may be quantified by the von Neumann entanglement 
entropy \cite{vNE1,vNE2,vNE3,vNE4}.
Denoting the reduced density matrix (RDM) of the subsystem by $\rho_{A}$,
the entanglement entropy is defined as
\begin{equation}
S=-\text{Tr}\rho_{A}\ln\rho_{A}.
\end{equation}
Alternatively the Rényi entanglement entropies 
\begin{equation}
S_{n}=\frac{1}{1-n}\ln\text{Tr}\rho_{A}^{n}\,,
\end{equation}
also provide bipartite entanglement measures in pure states and 
are related to the von Neumann one by taking the limit $n\rightarrow1$. 

The explicit idea of considering generally the internal structure if entanglement associated with symmetry is rather recent \cite{Greiner,gs-18,lr-14,SRENegativity}. 
In a symmetric state, the system's density matrix $\rho$ commutes with the conserved charge $\hat{Q}$ corresponding to the symmetry; 
if in addition $\hat{Q}_{A}$, the restriction of $\hat{Q}$ to this subsystem, satisfies
\begin{equation}
[\rho_{A},\hat{Q}_{A}]=0\,,\label{eq:Commutation}
\end{equation}
%where  is  and  $ \hat{Q}_{A}+\hat{Q}_{\bar{A}}=\hat{Q}$, 
then the RDM $\rho_{A}$ is block-diagonal with respect to the eigenspaces of $\hat{Q}_{A}$ and, consequently, the Rényi and
von Neumann entropies can be decomposed according to the symmetry sectors. 
Let us denote with $\mathcal{P}(q_{A})$ the projectors onto the eigenspace with eigenvalue $q_{A}$.  
The symmetry resolved partition functions can be defined as
\begin{equation}
\mathcal{Z}_{n}(q_{A})=\text{Tr}\left(\rho_{A}^{n}\mathcal{P}(q_{A})\right)\,,
\end{equation}
from which the symmetry resolved Rényi entropies  $S_{n}(q_{A})$
and the symmetry resolved von Neumann entropy $S(q_{A})$ can be naturally obtained as 
\begin{equation}
S_{n}(q_{A})=\frac{1}{1-n}\ln\left[\frac{\mathcal{Z}_{n}(q_{A})}{\mathcal{Z}_{1}^{n}(q_{A})}\right]\,,
\qquad
{\rm and}
\qquad 
S(q_{A})=-\frac{\partial}{\partial n}\left[\frac{\mathcal{Z}_{n}(q_{A})}{\mathcal{Z}_{1}^{n}(q_{A})}\right]_{n=1}\,,\label{eq:vNE}
\end{equation}
respectively. This way the total von Neumann entropy can be written
as \cite{ScPlusSf} 
\begin{equation}
S=\sum_{q_A}p(q_A)S(q_A)-\sum_{q_A}p(q_A)\ln p(q_A)=S^{c}+S^{f}\,,\label{eq:SfSc}
\end{equation}
where $p(q_A)=\mathcal{Z}_{1}(q_A)$ is the probability of finding $q_A$ as the outcome of a measurement of $\hat{Q}_{A}$. 
The contribution $S^{c}$ denotes the configurational entanglement entropy, which measures
the total entropy due to each charge sector (weighted with their probability)
\cite{Greiner,Wiseman} and $S^{f}$ denotes the fluctuation (or number) entanglement
entropy, which instead takes into account the entropy due to the fluctuations
of the value of the charge in the subsystem $A$ \cite{Greiner,FreeF3,MBL}.

The calculation of the symmetry resolved partition functions and entropies
is generally a difficult task; the usual way one proceeds includes the replica method and the computation of the charged moments \cite{gs-18}
\begin{equation}
Z_{n}(\alpha)=\text{Tr}\left(\rho_{A}^{n}e^{i\alpha\hat{Q}_{A}}\right).
\label{eq:Zn}
\end{equation}
Considering quantum field theories (QFTs) a natural way of computing
the Rényi entropies for integer $n$ is provided by the path-integral
formalism: $\text{Tr}\rho_{A}^{n}$ corresponds to the partition function
on an $n$-sheeted Riemann surface $\mathcal{R}_{n}$, which is obtained
by joining cyclically the $n$ sheets along the region $A$ \cite{cc-04,Replica,cc-09}.
It was recognised in \cite{gs-18} that the charged moments
(\ref{eq:Zn}) correspond, in the path integral language, to introducing
an Aharonov-Bohm flux on one of the sheets of $\mathcal{R}_{n}$.
An intuitive picture is given by imagining particles with a specific
charge eigenvalue moving from one level of $\mathcal{R}_{n}$ to the
other until they return to their original sheet \cite{gs-18};
if the charge within the subsystem is $q_A$,  the total acquired phase of a given particle is then $e^{i\alpha q_{A}}$
as given by the term $e^{i\alpha\hat{Q}_{A}}$ in Eq. (\ref{eq:Zn}). 
Focusing on $U(1)$ and $\mathbb{Z}_{N}$ discrete symmetries, the symmetry resolved
partition functions can then be computed by performing a continuous
or a discrete Fourier transform in the charge space as \cite{gs-18}
\begin{equation}
\mathcal{Z}_{n}(q_{A})=\text{Tr}\left(\rho_{A}^{n}\mathcal{P}(q_{A})\right)=\begin{cases}
\displaystyle
\int_{-\pi}^{\pi}\frac{\mathrm{d}\alpha}{2\text{\ensuremath{\pi}}}Z_{n}(\text{\ensuremath{\alpha}})e^{-i\alpha q_{A}}, &U(1)\;  \text{case}, \vspace{3mm}\\ 
\displaystyle
\frac{1}{N}\sum_{\alpha=0}^{N-1}Z_{n}(\text{\ensuremath{\alpha}})e^{-i\frac{2\pi\alpha q_{A}}{N}}, & \mathbb{Z_{N}}\text{ case ,}
\end{cases}\label{eq:CalZn}
\end{equation}
where $\alpha,q_{A}=0,\ldots,N-1$ in the $\mathbb{Z}_{N}$ case.
Symmetry resolved entropies have been studied in field theories including
conformal field theories (CFTs) \cite{gs-18,lr-14,Equipartitioning,SREQuench,crc-20}
and the free Dirac and complex boson field theories 
\cite{mdc-20b}, in lattice systems such as spin chains and hopping fermions/bosons
 \cite{brc-19,fg-20,FreeF1,FreeF2,mdc-20,lr-14,Equipartitioning,SREQuench,ccdm-20}
and also in the contexts of higher dimensional \cite{HigherDimFermions,mrc-20}, disordered systems \cite{MBL,trac-20}, 
and non-trivial topological phase \cite{Topology,Anyons}.
Finally we mention that charged moments like those in Eq. \eqref{eq:Zn} have been independently studied in field theoretical frameworks 
in several different circumstances \cite{cms-13,cnn-16,bym-13,d-16,ssr-17,srrc-19}.

In a path integral approach to quantum field theories (QFTs),  the computation of either $\text{Tr}\rho_{A}^{n}$ or
$\text{Tr}\rho_{A}^{n}e^{i\alpha\hat{Q}_{A}}$ can equivalently proceed for an $n$-copy QFT,
where specific boundary conditions are prescribed for the fields $\phi_{1},...,\phi_{n}$
corresponding to the different copies. 
Crucially, in 1+1 dimensional relativistic QFTs, there exist local fields in the $n$-copy theory that correspond
to the boundary conditions imposed on the fundamental fields in the path integral. 
These fields have been dubbed branch-point twist fields \cite{cc-04,ccd-08}.
The $n$th Rényi entropy of an arbitrary spatial subsystem (i.e. consisting also of disjoint intervals) is equivalent to a multi-point function
of the branch-point twist fields in an $n$-copy theory. 
Direct access to these fields is established in 2D CFT, where the scaling dimensions
of these fields are exactly known \cite{cc-04,k-87,dixon}.
These dimensions directly provide the scaling of two-points function, corresponding to a single interval  for a generic CFT \cite{cc-04}.
The behaviour of  four-point \cite{cct-09,h-10,cct-11,atc-11,rg-12,dei-18} and also higher functions \cite{ctt-14} of these twist fields are known for special CFTs.  
The main subject of this manuscript is however integrable quantum field theories (IQFTs). 
In these theories, the form factor (FF) bootstrap allows for the calculation of the matrix elements of the twist field \cite{ccd-08,cd-08,cd-09}.
Via the bootstrap, in principle, all matrix elements can be computed. However,
the correlation functions of the fields at large distances are usually well described by the first few members of the form factor series.
Such form factor bootstrap program has been used in IQFTs for the calculation of the entanglement entropy in many different 
situations \cite{cd-09b,cl-11,cdl-12,lcd-13,c-17,bc-16,cdds-18,clsv-19,clsv-20,bcd-15,bcd-16}.

The symmetry resolved entropies in CFT can be obtained by composite branch-point
twist fields in essentially the same way as the conventional entropies  \cite{gs-18}.
The only price to pay is the introduction of composite twist fields fusing the action of the replicas and of the flux of charge (see below for the precise definition). 
These new composite twist fields have been identified for Luttinger liquids \cite{gs-18}, for the $SU(2)_{k}$ Wess-Zumino-Witten models \cite{gs-18}, and for the Ising and
$\mathbb{Z}_{N}$ parafermion CFT \cite{fg-20}. 
Furthermore, the existence and applicability of such composite twist fields have been recently demonstrated for
the free massive Dirac and complex boson QFT too \cite{mdc-20b}.
These findings suggest that in perturbed QFTs (corresponding
to a relevant perturbation of a given CFT), the off critical version
of the composite twist field exists. We expect that in IQFTs their form factors
can be determined with the bootstrap program, similarly to the usual twist fields \cite{ccd-08,cd-08,cd-09}. 

This paper aims to initiate such a program for interacting
IQFTs. In particular, we introduce and discuss appropriate bootstrap
equations for the composite branch-point twist fields, find their
first few solutions and compute the long-distance leading behaviour of the symmetry resolved entropies
(similar twist fields have been introduced for non-unitary QFT \cite{bcd-15}, but in a completely different context and with different aims). 
For the sake of simplicity, here we consider the simplest integrable models, namely the Ising field theory, 
which is equivalent to a free Majorana fermion QFT, and the sinh-Gordon (ShG) model, which is a truly interacting QFT.
Both models possess the discrete $\mathbb{Z}_{2}$ symmetry. 
While from the point of view of IQFT techniques these models are indeed the simplest possible ones, the
resolution of their entanglement in terms of the $\mathbb{Z}_{2}$ symmetry requires a careful treatment because of the lack of a conserved density (\ref{eq:Commutation}). 
Integrable QFTs with continuous symmetry present many more technicalities because of their 
richer particle content and for the presence of  non-diagonal scattering. Their analysis is still on the way and will be eventually the subject of subsequent works.

The structure of this paper is as follows. In section \ref{sec:ConventionalTwistFields}
the FF approach for conventional branch-point twist fields is briefly
reviewed, focusing on the bootstrap equations and their solution for the Ising and ShG models. 
In section \ref{sec:Z2TwistFields}, we show how the bootstrap equations can be modified to obtain solutions
for the modified twist fields corresponding to a given symmetry resolution.
For the Ising and ShG models, the two-particle FFs of the $\mathbb{Z}_{2}$ twist fields are determined as well. 
Sections \ref{sec:Ising} and \ref{sec:ShG} are explicitly focused on Ising and ShG models respectively, 
reporting also $\Delta$-theorem \cite{delta_theorem} checks of the obtained form factors;
for the Ising model the even particle-number FFs are expressed in terms of a Pfaffian involving the two-particle matrix elements. 
Section \ref{sec:gen} reports general results for $\mathbb{Z}_{2}$ symmetry resolved entropies that can be deduced from the IQFT structure. 
The leading and sub-leading contributions of the symmetry resolved entanglement are explicitly calculated in section
\ref{sec:CalculatingEntropies} for the paramagnetic ground state of the Ising model. 
%Our results are cross-checked against lattice results after computing their continuum limit.
%We show that there is equipartition of entanglement for both Ising and ShG models, up to $\mathcal{O}(\varepsilon^{\frac{1}{4}}\ln\varepsilon)$ corrections, 
%where $\varepsilon$ is a UV regulator. 
We conclude in section \ref{sec:Conclusions}, which is followed by the appendices containing the determination of the vacuum expectation value (VEV) of the
Ising $\mathbb{Z}_{2}$ branch-point twist field (appendix \ref{sec:Appendix-A-VEV}) and some auxiliary calculations.

\section{Form factors of the branch-point twist fields in integrable models\label{sec:ConventionalTwistFields}}

Before presenting our results and discussing the determination of
the form factors of modified branch-point twist fields, it is instructive
to give a brief overview of some basic ingredients of IQFTs and in
particular on form factors of the conventional branch-point twist
fields. Here we mostly follow the logic of Ref. \cite{ccd-08}
and present some of its results with an emphasis on the bootstrap
equation.

Form factors (FF) are matrix elements of (semi-)local operators $O(x,t)$
between the vacuum and asymptotic states, i.e., 
\begin{equation}
F_{\alpha_{1},\ldots,\alpha_{n}}^{O}(\vartheta_{1},\ldots,\vartheta_{n})=\langle0|O(0,0)|\vartheta_{1},\ldots\vartheta_{n}\rangle_{\alpha_{1},\ldots,\alpha_{n}}.\label{eq:FF}
\end{equation}
In massive field theories, the asymptotic states correspond to multi-particle
excitations, with dispersion relation $(E,p)=(m_{\alpha_{i}}\cosh\vartheta,m_{\alpha_{i}}\sinh\vartheta)$,
where $\alpha_{i}$ indicates the particle species. In such models,
any multi-particle state can be constructed from vacuum state by means
of the particle creation operators $A_{\alpha_{i}}^{\dagger}(\vartheta)$
by 
\begin{equation}
|\vartheta_{1},\vartheta_{2},...,\vartheta_{n}\rangle=A_{\alpha_{1}}^{\dagger}(\vartheta_{1})A_{\alpha_{2}}^{\dagger}(\vartheta_{2})\ldots.A_{\alpha_n}^{\dagger}(\vartheta_{n})|0\rangle\:,\label{eq:basis}
\end{equation}
where the operator $A_{\alpha_{i}}^{\dagger}(\vartheta)$ creates
a particle of species $\alpha_{i}$ with rapidity $\vartheta$ and
$|0\rangle$ is the vacuum state of the theory. In an IQFT with factorized
scattering, the creation and annihilation operators $A_{\alpha_{i}}^{\dagger}(\vartheta)$
and $A_{\alpha_{i}}(\vartheta)$ satisfy the Zamolodchikov-Faddeev
(ZF) algebra 
\begin{eqnarray}
A_{\alpha_{i}}^{\dagger}(\vartheta_{i})A_{\alpha_{j}}^{\dagger}(\vartheta_{j}) & = & S_{\alpha_{i},\alpha_{j}}(\vartheta_{i}-\vartheta_{j})A_{\alpha_{j}}^{\dagger}(\vartheta_{j})A_{\alpha_{i}}^{\dagger}(\vartheta_{i})\:,\nonumber \\
A_{\alpha_{i}}(\vartheta_{i})A_{\alpha_{j}}(\vartheta_{j}) & = & S_{\alpha_{i},\alpha_{j}}(\vartheta_{i}-\vartheta_{j})A_{\alpha_{j}}(\vartheta_{j})A_{\alpha_{i}}(\vartheta_{i})\:,\nonumber \\
A_{\alpha_{i}}(\vartheta_{i})A_{\alpha_{j}}^{\dagger}(\vartheta_{j}) & = & S_{\alpha_{i},\alpha_{j}}(\vartheta_{j}-\vartheta_{i})A_{\alpha_{j}}^{\dagger}(\vartheta_{j})A_{\alpha_{i}}(\vartheta_{i})+\delta_{\alpha_{i},\alpha_{j}}2\pi\delta(\vartheta_{i}-\vartheta_{j}),\label{eq:ZF}
\end{eqnarray}
where $S_{\alpha_{i},\alpha_{j}}(\vartheta_{i}-\vartheta_{j})$ are
the two-particle S-matrices of the theory.

Our primary interest now is an $n$-copy IQFT and the corresponding
branch-point twist fields. For simplicity we assume that there is
only one particle in the original theory. Then the scattering between
the particles of different and of the same copies is described by
\begin{equation}
\begin{split}S_{i,j}(\vartheta) & =1,\;\qquad i,j=1,...,n\text{ and }i\neq j,\\
\quad\quad\quad S_{i,i}(\vartheta) & =S(\vartheta),\;\;\;i=1,..,n,
\end{split}
\end{equation}
and the branch-point twist fields are related to the symmetry $\sigma\Psi_{i}=\Psi_{i+1}$,
where $n+i\equiv i$. The insertion of a twist field $\mathcal{T}$
(or ${\cal T}_{n}$) in a correlation function can be summarised as
\begin{equation}
\begin{split}\Psi_{i}(y)\mathcal{T}(x)=\mathcal{T}(x)\Psi_{i+1}(y) & \qquad x>y,\\
\Psi_{i}(y)\mathcal{T}(x)=\mathcal{T}(x)\Psi_{i}(y) & \qquad x<y,
\end{split}
\end{equation}
and we can also define $\tilde{\mathcal{T}}$, whose action is 
\begin{equation}
\begin{split}\Psi_{i}(y)\mathcal{\tilde{\mathcal{T}}}(x)=\mathcal{\tilde{\mathcal{T}}}(x)\Psi_{i-1}(y) & \qquad x{>y},\\
\Psi_{i}(y)\mathcal{\tilde{\mathcal{T}}}(x)=\tilde{\mathcal{T}}(x)\Psi_{i}(y) & \qquad x<y.
\end{split}
\end{equation}

The form factors of the branch-point twist fields satisfy the following
relations, which are simple modifications of the form factor bootstrap
equations \cite{bergkarowski,kirillovsmirnov,FFAxioms} 
\begin{eqnarray}
 &  & F_{k}^{\mathcal{T}|...\mu_{i},\mu_{i+1}...}(\ldots\vartheta_{i},\vartheta_{i+1},\ldots)=S_{\mu_{i},\mu_{i+1}}(\vartheta_{i,i+1})F_{k}^{\mathcal{T}|...\mu_{i+1},\mu_{i}...}(\ldots\vartheta_{i+1},\vartheta_{i},\ldots),\label{eq:FFAxiom1}\\
 &  & F_{k}^{\mathcal{T}|\mu_{1},\mu_{2},...,\mu_{k}}(\vartheta_{1}+2\pi i,\vartheta_{2},\ldots,\vartheta_{k})=F_{k}^{\mathcal{T}|\mu_{2},...,\mu_{k},\hat{\mu}_{1}}(\vartheta_{2},\ldots,\vartheta_{n},\vartheta_{1}),\label{eq:FFAxiom2}\\
 &  & -i\underset{\vartheta_{0}'=\vartheta_{0}+i\pi}{{\rm Res}}F_{k+2}^{\mathcal{T}|\mu,\mu,\mu_{1},\mu_{2},...,\mu_{k}}(\vartheta_{0}',\vartheta_{0},\vartheta_{1},\vartheta_{2},\ldots,\vartheta_{k})=F_{k}^{\mathcal{T}|\mu_{1},\mu_{2},...,\mu_{k}}(\vartheta_{1},\vartheta_{2},\ldots,\vartheta_{k}),\label{eq:FFAxiom3}\\
 &  & -i\underset{\vartheta_{0}'=\vartheta_{0}+i\pi}{{\rm Res}}F_{k+2}^{\mathcal{T}|\mu,\hat{\mu},\mu_{1},\mu_{2},...,\mu_{k}}(\vartheta_{0}',\vartheta_{0},\vartheta_{1},\vartheta_{2},\ldots,\vartheta_{k})=-\prod_{i=1}^{k}S_{\hat{\mu},\mu_{i}}(\vartheta_{0i})F_{k}^{\mathcal{T}|\mu_{1},\mu_{2},...,\mu_{k}}(\vartheta_{1},\vartheta_{2},\ldots,\vartheta_{k}),\nonumber 
\end{eqnarray}
where $\mu$ refers to the replica index of the particle,
$\vartheta_{ij}=\vartheta_{i}-\vartheta_{j}$ and $\hat{\mu}=\mu+1$.
In addition relativistic invariance implies 
\begin{equation}
F_{k}^{\mathcal{T}|\mu_{1},\mu_{2},...,\mu_{k}}(\vartheta_{1}+\Lambda,\ldots,\vartheta_{k}+\Lambda)=e^{s\Lambda}F_{k}^{\mathcal{T}|\mu_{1},\mu_{2},...,\mu_{k}}(\vartheta_{1},\ldots,\vartheta_{k}),\label{eq:RelInv}
\end{equation}
where $s$ is the Lorentz spin of the operator, which
is zero for the branch-point twist fields. As the theories we consider
in this paper have no bound states, Eqs. (\ref{eq:FFAxiom1})-(\ref{eq:FFAxiom3})
and (\ref{eq:RelInv}) give all the constraints for form factors of
the twist fields.

As usual in this context, the so-called minimal form factor $F_{\text{min}}^{\mathcal{T}|j,k}(\vartheta,n)$
is defined as the solution of the first two equations, Eqs. (\ref{eq:FFAxiom1}) and (\ref{eq:FFAxiom2}). That is, the minimal form factor satisfies
\begin{equation}
F_{\text{min}}^{\mathcal{T}|k,j}(\vartheta,n)=F_{\text{min}}^{\mathcal{T}|j,k}(-\vartheta,n)S_{k,j}(\vartheta)=F_{\text{min}}^{\mathcal{T}|j,k+1}(2\pi i-\vartheta,n)\,.
\end{equation}
It is then easy to show that 
\begin{equation}
\ \begin{split}F_{\text{min}}^{\mathcal{T}|i,i+k}(\vartheta,n)= & F_{\text{min}}^{\mathcal{T}|j,j+k}(\vartheta,n)\quad\forall i,j,k\\
F_{\text{min}}^{\mathcal{T}|1,j}(\vartheta,n)= & F_{\text{min}}^{\mathcal{T}|1,1}(2\pi i(j-1)-\vartheta,n)\quad\forall j\neq1\,,
\end{split}
\label{eq:FMinAxioms}
\end{equation}
from which it follows that 
\begin{equation}
F_{\text{min}}^{\mathcal{T}|j,k}(\vartheta,n)=\begin{cases}
F_{\text{min}}^{\mathcal{T}|1,1}(2\pi i(k-j)-\vartheta,n) & \text{ if }k>j,\\
F_{\text{min}}^{\mathcal{T}|1,1}(2\pi i(j-k)+\vartheta,n) & \text{otherwise},
\end{cases}
\end{equation}
and hence the only independent quantity is $F_{\text{min}}^{\mathcal{T}|1,1}(\vartheta,n)$.
We can use Eq. (\ref{eq:FMinAxioms}) to determine it, writing 
\begin{equation}
F_{\text{min}}^{\mathcal{T}|1,1}(\vartheta,n)=F_{\text{min}}^{\mathcal{T}|1,1}(-\vartheta,n)S(\vartheta)=F_{\text{min}}^{\mathcal{T}|1,1}(-\vartheta+2\pi in,n)\,.
\end{equation}
The solution of the last equation is easily obtained by noticing that
if it exists a function $f_{11}(\vartheta)$ satisfying 
\begin{equation}
f_{11}(\vartheta)=f_{11}(-\vartheta)S(n\vartheta)=f_{11}(-\vartheta+2\pi i)\,,\label{f11axiom}
\end{equation}
then 
\begin{equation}
F_{\text{min}}^{\mathcal{T}|1,1}(\vartheta,n)=f_{11}(\vartheta/n)\,.
\end{equation}
Eq. (\ref{f11axiom}) is, nevertheless, the standard equation for minimal form factors of conventional local operators, but with an S-matrix $S(n\vartheta)$ instead of $S(\vartheta)$. 
When $S(\vartheta)$ can be parametrised as 
\begin{equation}
S(\vartheta)=\exp\left[\int_{0}^{\infty}\frac{\mathrm{d}t}{t}g(t)\sinh\frac{t\vartheta}{i\pi}\right],\label{Svsg}
\end{equation}
with some function $g(t)$, the minimal FF is 
\begin{equation}
f_{11}(\vartheta)=\mathcal{N}\exp\left[\int_{0}^{\infty}\frac{\mathrm{d}t}{t}\frac{g(t)}{\sinh nt}\sin^{2}\left(\frac{itn}{2}\left(1+\frac{i\vartheta}{\pi}\right)\right)\right]\,,\label{eq:f11}
\end{equation}
where the normalisation $\mathcal{N}$ ensures that $f_{11}(\pm\infty)=1$
and thus 
\begin{equation}
F_{\text{min}}^{\mathcal{T}|1,1}(\vartheta,n)=\mathcal{N}\exp\left[\int_{0}^{\infty}\frac{\mathrm{d}t}{t\sinh nt}g(t)\sin^{2}\left(\frac{it}{2}\left(n+\frac{i\vartheta}{\pi}\right)\right)\right]\,.
\end{equation}

The minimal form factors are very useful to obtain all form factors
with particle number $k\geq2$ as they can be used as building blocks,
hence simplifying the solution of the bootstrap equations. The zero
and one-particle form factors have to be determined by other means.
The most important quantities are usually two-particle form factors.
It can be verified that the two-particle form factors for the branch-point
twist field, satisfying also the kinematic poles axioms, read \cite{ccd-08}
\begin{equation}
F_{2}^{\mathcal{T}|j,k}(\vartheta,n)=\frac{\langle\mathcal{T}_{n}\rangle\sin\frac{\pi}{n}}{2n\sinh\left(\frac{i\pi\left(2(j-k)-1\right)+\vartheta}{2n}\right)\sinh\left(\frac{i\pi\left(2(k-j)-1\right)-\vartheta}{2n}\right)}\frac{F_{\text{min}}^{\mathcal{T}|j,k}(\vartheta,n)}{F_{\text{min}}^{\mathcal{T}|1,1}(i\pi,n)}\,,\label{eq:F2TwistField}
\end{equation}
where $\langle\mathcal{T}_{n}\rangle=F_{0}^{\mathcal{T}}$ is the
vacuum expectation value (VEV) of ${\cal T}$. Furthermore, relativistic
invariance implies that $F_{2}^{\mathcal{T}|j,k}(\vartheta_{1},\vartheta_{2},n)$
depends only on the rapidity difference $\vartheta_{1}-\vartheta_{2},$
justifying writing $F_{2}^{\mathcal{T}|j,k}(\vartheta_{1}-\vartheta_{2},n)$
or merely $F_{2}^{\mathcal{T}|j,k}(\vartheta,n)$. It straightforward
to show that for $\hat{\mathcal{T}}$ we have 
\begin{equation}
F_{2}^{\mathcal{T}|j,k}(\vartheta,n)=F_{2}^{\hat{\mathcal{T}}|n-j,n-k}(\vartheta,n)\,.\label{FFtil}
\end{equation}

\subsection{Branch-point twist field form factors in the Ising model}

The Ising field theory is surely the easiest integrable field theory.
It has one massive particle (a free Majorana fermion) and the simple S-matrix 
\begin{equation}
S_{\text{Ising}}(\vartheta)=-1,
\end{equation}
and consequently 
\begin{equation}
F_{\text{min}}^{\mathcal{T}|1,1}(\vartheta,n)=-i\sinh\frac{\vartheta}{2n}\,.
\end{equation}
For this model, it has been shown that the FFs of the branch-point
twist fields are only non-vanishing for even particle number \cite{ccd-08,cd-09}.
Moreover, the FFs for any even $n$ can be written as a Pfaffain of
the two-particle FF \cite{cd-09b}.

\subsection{Branch-point twist field form factors in the sinh-Gordon model}

The sinh-Gordon model, with Euclidean action 
\begin{equation}
{\cal S}=\int{\rm d}^{2}x\left\{ \frac{1}{2}\left[\partial\phi(x)\right]^{2}+\frac{\mu^{2}}{g^{2}}:\!\cosh\left[g\phi(x)\right]\!:\right\} ,\label{hamiltonian}
\end{equation}
is arguably the simplest interacting integrable relativistic QFT and
for this reason it is often taken as a reference point and has been
the subject of an intense research activity since many decades, see,
e.g., \cite{m-book,AFZ,FMS,SinhGordonFF,ShG-Ising,ns-13,bpc-16,d-18,klm-20}.
Furthermore, it recently became also experimentally relevant because
its non-relativistic limit is the Lieb-Liniger Bose gas \cite{ll-63},
a paradigmatic model for 1D ultracold gases \cite{rev1d}. This limit,
joined with the FF program, allowed for the calculation of many quantities
that were too difficult, or even impossible, by other means \cite{bdm-16,bp-18,bpc-18,kmt-10,kmt-09,kmp-10}.

The spectrum of the model consists of multi-particle states of
a single massive bosonic particle. The two-particle S-matrix is given
by \cite{AFZ} 
\begin{equation}
S_{\text{ShG}}(\theta)=\frac{\tanh\frac{1}{2}\left(\vartheta-i\frac{\pi B}{2}\right)}{\ \tanh\frac{1}{2}\left(\vartheta+i\frac{\pi B}{2}\right)}\:,\label{eq:SinhGordonS}
\end{equation}
where $B$ is defined as 
\begin{equation}
B(g)\,=\,\frac{2g^{2}}{8\pi+g^{2}}\:.
\end{equation}
For the ShG model, the solutions of the system (\ref{eq:FFAxiom1})-(\ref{eq:RelInv})
have been constructed in \cite{ShGFFZamo,FMS,SinhGordonFF}.

The function $g(t)$ entering in the parametrisation of the S-matrix
\eqref{Svsg} can be identified with 
\begin{equation}
g(t)=\frac{8\sinh\left(\frac{tB}{2}\right)\sinh\left(\frac{t}{2}\left(1-\frac{B}{2}\right)\right)\sinh\left(\frac{t}{2}\right)}{\sinh t}\,,
\end{equation}
from which 
\begin{equation}
F_{\text{min,ShG}}^{\mathcal{T}|1,1}(\vartheta,n)=\exp\left[-2\int_{0}^{\infty}\frac{\mathrm{d}t}{t}\frac{\sinh\left(\frac{tB}{4}\right)\sinh\left(\frac{t}{4}\left(2-B\right)\right)}{\sinh\left(nt\right)\cosh\left(\frac{t}{2}\right)}\cosh\left(t\left(n+\frac{i\vartheta}{\pi}\right)\right)\right]\,.
\end{equation}
It is possible to write down an alternative representation of $F_{\text{min,ShG}}^{\mathcal{T}|1,1}(\vartheta,n)$ in terms of infinite products \cite{ccd-08}. 
For and efficient numerical computation the following  mixed representation is more useful
\begin{equation}
\begin{split}F_{\text{min,ShG}}^{\mathcal{T}|1,1}(\vartheta,n)= & \prod_{k=0}^{m}\left[\frac{\Gamma\left(\frac{2k+2n+\frac{i\theta}{\pi}+2}{2n}\right)\Gamma\left(\frac{B+4k+2n-2\left(n+\frac{i\theta}{\pi}\right)}{4n}\right)\Gamma\left(\frac{2-B+4k+2n-2\left(n+\frac{i\theta}{\pi}\right)}{4n}\right)}{\Gamma\left(\frac{2k+2n+\frac{i\theta}{\pi}}{2n}\right)\Gamma\left(\frac{B+4k+2n-2\left(n+\frac{i\theta}{\pi}\right)+2}{4n}\right)\Gamma\left(\frac{4-B+4k+2n-2\left(n+\frac{i\theta}{\pi}\right)}{4n}\right)}\times\right.\\
 & \times\left.\frac{\Gamma\left(\frac{2k-\frac{i\theta}{\pi}+2}{2n}\right)\Gamma\left(\frac{2-B+4k+2n+2\left(n+\frac{i\theta}{\pi}\right)}{4n}\right)\Gamma\left(\frac{B+4k+2n+2\left(n+\frac{i\theta}{\pi}\right)}{4n}\right)}{\Gamma\left(\frac{2k-\frac{i\theta}{\pi}}{2n}\right)\Gamma\left(\frac{4-B+4k+2n+2\left(n+\frac{i\theta}{\pi}\right)}{4n}\right)\Gamma\left(\frac{2+B+4k+2n+2\left(n+\frac{i\theta}{\pi}\right)}{4n}\right)}\right]\times\\
 & \times\exp\left[-4\int_{0}^{\infty}\frac{\mathrm{d}t}{t}\frac{\sinh\left(\frac{Bt}{4}\right)\sinh\left(\frac{t}{4}(2-B)\right)\cosh\left(t\left(n+\frac{i\theta}{\pi}\right)\right)e^{-\frac{t}{2}}e^{-t\left(2m+2\right)}}{(e^{-t}+1)\sinh(nt)}\right]\,.
\end{split}
\end{equation}
Similarly to the Ising model, the FFs of the ShG branch-point twist fields are only non-vanishing for even particle number \cite{ccd-08,cd-09}.

A very important relation between the ShG and Ising models is that the S-matrix and certain form factors of the ShG theory collapse to
that of the Ising model, when the limit $B=1+i\frac{2}{\pi}\Theta_{0}$ with $\Theta_{0}\rightarrow\infty$ is taken \cite{ShG-Ising}. 
It can be checked that both $F_{\text{min,ShG}}^{\mathcal{T}|1,1}(\vartheta,n)$ and $F_{2,\text{ShG}}^{\mathcal{T}|j,k}(\vartheta,n)$ in this limit
collapse to the corresponding quantities in Ising model.
This limit will be an important guide for the case of the composite twist fields discussed below.

\section{Form factors of the composite branch-point twist fields for $\mathbb{Z}_{2}$ symmetry in integrable models}
\label{sec:Z2TwistFields}

After the introduction of the bootstrap equations for the FFs of the
branch-point twist field, we now show how these equations can be naturally
modified to obtain the corresponding quantities of the composite twist
fields. At this point, of course, the existence of such fields is not
strictly justified, therefore the formal solutions of the modified
bootstrap equations will be subject to subsequent cross-checks.

To achieve our goal, first of all, we define the semi-local (or mutual
locality) index $e^{2\pi i\gamma}$ of an operator $O$ with respect
to the interpolating field $\phi$ via the condition 
\begin{equation}
O(x,t)\phi(y,t')=e^{i2\pi\gamma}\phi(y,t')O(x,t),\label{eq:LocalityDef}
\end{equation}
for space-like separated space-time points. Local operators correspond
to $e^{i2\pi\gamma}=1$, while fields with $e^{i2\pi\gamma}\neq1$
are called semi-local. It is natural to assume that the phase $e^{i\alpha}$
corresponding to the flux can be related with the mutual locality
index appearing in the bootstrap equation. 
This assumption can be based on the intuitive picture associated with the insertion of the Aharonov-Bohm flux on one of the Riemann sheets. 
In this picture, the flux is carried by the particles of the theory, but Eq. (\ref{eq:LocalityDef}) is just an equivalent rephrasing of this idea because 
the interpolating field is associated with creating/annihilating particles.

To be more precise about the connection between $e^{i2\pi\gamma}$
and $e^{i\alpha}$, let us consider briefly a $U(1)$ symmetry for
which $\alpha$ is a continuous parameter. From the point of view
of the bootstrap equations, it is more convenient not to favour any
of the Riemann sheets by adding the flux to it, but rather to divide
the flux and introducing it on all sheets. This procedure corresponds
to add a phase $e^{i\alpha/n}$ on each sheet and therefore the locality
factor $e^{i2\pi\gamma}$ and $e^{i\alpha/n}$ must be equal. The
further elaboration of the $U(1)$ symmetry will be the subject of
a subsequent work because, in this case, the particle content of the
IQFT is richer and allows also for non-diagonal scattering leading
to more complicated form factors. Here, we focus on the simpler, yet
not trivial, analysis of the $\mathbb{Z}_{2}$ symmetry in models
with only one particle species.

However, for the $\mathbb{Z}_{2}$ symmetry (and more generally for
discrete symmetries) there are two subtleties that we cannot avoid
mentioning. The first one is rather fundamental: for discrete symmetries
Noether's theorem does not guarantee the existence of a conserved
density, hence it is not a priori obvious if the reduced density matrix
commutes with the symmetry operator. This problem will be discussed
in the following sections for the specific cases of the Ising and
ShG QFT. The other issue is that the phase is $e^{i\pi}=-1$ cannot
be divided as $e^{i\pi/n}$ among the various sheets, because $e^{i\pi/n}$
no longer corresponds to the $\mathbb{Z}_{2}$ symmetry of interest.
This latter difficulty can be easily overcome by introducing the flux
corresponding to the phase $e^{i\pi}=-1$ on all sheets. This step
is legitimate if the number of sheets $n$ is odd, as the overall
phase acquired by a hypothetical particle winded through all sheets
is still $(-1)^{n}=-1$. Our argument implies that the composite branch-point
twist fields associated with the $\mathbb{Z}_{2}$ symmetry in the
Ising and ShG models is a semi-local operator with respect to the
fundamental field, with locality index $e^{2\pi i\gamma}=-1$. Specialising
the bootstrap equations of a generic semi-local twist field 
\begin{eqnarray}
&& F_{k}^{\mathcal{T}|...\mu_{i},\mu_{i+1}...}(\ldots\vartheta_{i},\vartheta_{i+1},\ldots)=S_{\mu_{i},\mu_{i+1}}(\vartheta_{i,i+1})F_{k}^{\mathcal{T}|...\mu_{i+1},\mu_{i}...}(\ldots\vartheta_{i+1},\vartheta_{i},\ldots),\label{eq:FFAxiom1SemiLocalGeneral}\\
&& F_{k}^{\mathcal{T}|\mu_{1},\mu_{2},...,\mu_{k}}(\vartheta_{1}+2\pi i,\vartheta_{2},\ldots,\vartheta_{k})=e^{2\pi i\gamma}F_{k}^{\mathcal{T}|\mu_{2},...,\mu_{k},\hat{\mu}_{1}}(\vartheta_{2},\ldots,\vartheta_{n},\vartheta_{1}),\label{eq:FFAxiom2SemiLocalGeneral}\\
&& -i\underset{\vartheta'=\vartheta+i\pi}{\rm Res}F_{k+2}^{\mathcal{T}|\mu,\mu,\mu_{1},\mu_{2},...,\mu_{k}}(\vartheta_{0}',\vartheta_{0},\vartheta_{1},\vartheta_{2},\ldots,\vartheta_{k})=F_{k}^{\mathcal{T}|\mu_{1},\mu_{2},...,\mu_{k}}(\vartheta_{1},\vartheta_{2},\ldots,\vartheta_{k}),\label{eq:FFAxiom3SemiLocalGeneral}\\
&& -i\underset{\vartheta'=\vartheta+i\pi}{\rm Res}F_{k+2}^{\mathcal{T}|\mu,\hat{\mu},\mu_{1},...,\mu_{k}}(\vartheta_{0}',\vartheta_{0},\vartheta_{1},\vartheta_{2},\ldots,\vartheta_{k})=-e^{2\pi i\gamma}\prod S_{\hat{\mu},\mu_{i}}(\vartheta_{0i})F_{k}^{\mathcal{T}|\mu_{1},...,\mu_{k}}(\vartheta_{1},\vartheta_{2},\ldots,\vartheta_{k}),\nonumber 
\end{eqnarray}
to the $\mathbb{Z}_{2}$ case, we have 
\begin{eqnarray}
 &  & F_{k}^{\mathcal{T}^{D}|...\mu_{i},\mu_{i+1}...}(\ldots\vartheta_{i},\vartheta_{i+1},\ldots)=S_{\mu_{i},\mu_{i+1}}(\vartheta_{i,i+1})F_{k}^{\mathcal{T}^{D}|...\mu_{i+1},\mu_{i}...}(\ldots\vartheta_{i+1},\vartheta_{i},\ldots),\label{eq:FFAxiom1SemiLocalZ2}\\
 &  & F_{k}^{\mathcal{T}^{D}|\mu_{1},\mu_{2},...,\mu_{k}}(\vartheta_{1}+2\pi i,\vartheta_{2},\ldots,\vartheta_{k})=-F_{k}^{\mathcal{T}^{D}|\mu_{2},...,\mu_{k},\hat{\mu}_{1}}(\vartheta_{2},\ldots,\vartheta_{n},\vartheta_{1}),\label{eq:FFAxiom2SemiLocalZ2}\\
 &  & -i\underset{\vartheta_{0}'=\vartheta_{0}+i\pi}{\rm Res}F_{k+2}^{\mathcal{T}^{D}|\mu,\mu,\mu_{1},\mu_{2},...,\mu_{k}}(\vartheta_{0}',\vartheta_{0},\vartheta_{1},\vartheta_{2},\ldots,\vartheta_{k})=F_{k}^{\mathcal{T}^{D}|\mu_{1},\mu_{2},...,\mu_{k}}(\vartheta_{1},\vartheta_{2},\ldots,\vartheta_{k}),\label{eq:FFAxiom3SemiLocalZ2}\\
 &  & -i\underset{\vartheta_{0}'=\vartheta_{0}+i\pi}{\rm Res}F_{k+2}^{\mathcal{T}^{D}|\mu,\hat{\mu},\mu_{1},...,\mu_{k}}(\vartheta_{0}',\vartheta_{0},\vartheta_{1},\vartheta_{2},\ldots,\vartheta_{k})=\prod S_{\hat{\mu},\mu_{i}}(\vartheta_{0i})F_{k}^{\mathcal{T}^{D}|\mu_{1},...,\mu_{k}}(\vartheta_{1},\vartheta_{2},\ldots,\vartheta_{k}),\nonumber 
\end{eqnarray}
where $\mathcal{T}^{D}$ denotes the composite branch-point twist
field associated with the $\mathbb{Z}_{2}$ symmetry. Having obtained
the defining equations, following the logic of section \ref{sec:ConventionalTwistFields},
we can write 
\begin{equation}
F_{\text{min}}^{\mathcal{T}^{D}|k,j}(\vartheta,n)=F_{\text{min}}^{\mathcal{T}^{D}|j,k}(-\vartheta,n)S_{k,j}(\vartheta)=-F_{\text{min}}^{\mathcal{T}^{D}|j,k+1}(2\pi i-\vartheta,n)\,,
\end{equation}
for the minimal form factor $F_{\text{min}}^{\mathcal{T}^{D}}$ of the composite twist field ${\mathcal{T}^{D}}$. From this we find 
\begin{equation}
\begin{split}  
F_{\text{min}}^{\mathcal{T}^{D}|i,i+k}(\vartheta,n)=F_{\text{min}}^{\mathcal{T}^{D}|j,j+k}(\vartheta,n)&\ \quad\forall i,j,k,\\
  F_{\text{min}}^{\mathcal{T}^{D}1,j}(\vartheta,n)=(-1)^{(j-1)}F_{\text{min}}^{\mathcal{T}^{D}|1,1}(2\pi i(j-1)-\vartheta,n)&\ \quad\forall j\neq1,
\end{split}
\label{eq:FMinAxiomsSemiLocalZ2}
\end{equation}
and finally we get
\begin{equation}
F_{\text{min}}^{\mathcal{T}^{D}|j,k}(\vartheta,n)=(-1)^{(k-j)}\begin{cases}
F_{\text{min}}^{\mathcal{T}^{D}|1,1}(2\pi i(k-j)-\vartheta,n) & \text{ if }k>j,\\
F_{\text{min}}^{\mathcal{T}^{D}|1,1}(2\pi i(j-k)+\vartheta,n) & \text{otherwise.}
\end{cases}\label{eq:FDMinBasicProp}
\end{equation}
Akin to the previous case, the only independent quantity is $F_{\text{min}}^{\mathcal{T}^{D}|1,1}(\vartheta,n)$.
We exploit Eq. (\ref{eq:FMinAxiomsSemiLocalZ2}) to write for odd $n$
\begin{equation}
F_{\text{min}}^{\mathcal{T}^{D}|1,1}(\vartheta,n)=F_{\text{min}}^{\mathcal{T}^{D}|1,1}(-\vartheta,n)S(\vartheta)=-F_{\text{min}}^{\mathcal{T}^{D}|1,1}(-\vartheta+2\pi in,n)\,.
\end{equation}
For even $n$ the above equation is equal to that of $F_{\text{min}}^{\mathcal{T}|1,1}(\vartheta,n)$,
but our analysis is valid only for odd $n$. The solution of $F_{\text{min}}^{\mathcal{T}^{D}|1,1}$
can be obtained by introducing $f_{11}^{D}(\vartheta)$ as 
\begin{equation}
F_{\text{min}}^{\mathcal{T}^{D}|1,1}(\vartheta,n)=f_{11}^{D}(\vartheta/n)\,,
\end{equation}
that satisfies 
\begin{equation}
f_{11}^{D}(\vartheta)=f_{11}^{D}(-\vartheta)S(n\vartheta)=-f_{11}^{D}(-\vartheta+2\pi i)\,.\label{f11axiomSemiLocalZ2}
\end{equation}
Luckily, $f_{11}^{D}$ can be easily obtained from $f_{11}$ by multiplying
the latter by an appropriately chosen CDD factor, $f_{\text{CDD}}$.
Such a factor must obey 
\begin{equation}
f_{\text{CDD}}(\vartheta)=f_{\text{CDD}}(-\vartheta)=-f_{\text{CDD}}(-\vartheta+2\pi i),\label{eq:fCDDAxiom}
\end{equation}
guaranteeing that $f_{11}^{D}(\vartheta)=f_{\text{CDD}}(\vartheta)f_{11}(\vartheta)$
satisfies Eq. (\ref{f11axiomSemiLocalZ2}). The correct choice for
$f_{\text{CDD}}$ turns out to be 
\begin{equation}
f_{\text{CDD}}(\vartheta)=2\cosh\frac{\vartheta}{2}\,.\label{eq:fCDDSemiLocalZ2}
\end{equation}
It is easy to check that the ansatz (\ref{eq:fCDDSemiLocalZ2}) satisfies Eq. (\ref{eq:fCDDAxiom}), but it is not entirely trivial that there
is no further ambiguity for the CDD factor and that Eq. (\ref{eq:fCDDSemiLocalZ2})
is the correct choice for both the Ising and ShG models. Some tests
of this statement are carried out in the next sections for both models
by studying the limit $n\rightarrow1$ of the form factors $F_{2}^{\mathcal{T}^{D}|j,k}$
and by exploiting the $\Delta$-theorem.

Putting the various pieces together, the minimal form factor of the
composite twist field is 
\begin{equation}
F_{\text{min}}^{\mathcal{T}^{D}|1,1}(\vartheta,n)=2\cosh\Big(\frac{\vartheta}{2n}\Big)F_{\text{min}}^{\mathcal{T}|1,1}(\vartheta,n)\,.
\end{equation}
Given this minimal form factor, it is easy to show that Eq. (\ref{eq:F2TwistField})
for two-particle form factors is still valid, i.e. 
\begin{equation}
F_{2}^{\mathcal{T}^{D}|j,k}(\vartheta,n)=\frac{\langle\mathcal{T}_{n}^{D}\rangle\sin\frac{\pi}{n}}{2n\sinh\left(\frac{i\pi\left(2(j-k)-1\right)+\vartheta}{2n}\right)\sinh\left(\frac{i\pi\left(2(k-j)-1\right)-\vartheta}{2n}\right)}\frac{F_{\text{min}}^{\mathcal{T}^{D}|j,k}(\vartheta,n)}{F_{\text{min}}^{\mathcal{T}^{D}|1,1}(i\pi,n)}\,,\label{eq:F2TwistFieldSemiLocalZ2}
\end{equation}
for odd $n$, where $\langle\mathcal{T}_{n}^{D}\rangle=F_{0}^{\mathcal{T}^{D}}$
is the vacuum expectation value of $\mathcal{T}^{D}$. Again, relativistic
invariance implies that $F_{2}^{\mathcal{T}^{D}|j,k}(\vartheta_{1},\vartheta_{2},n)$
depends only on the rapidity difference $\vartheta_{1}-\vartheta_{2},$
thus we can write $F_{2}^{\mathcal{T}^{D}|j,k}(\vartheta,n)$. 
It is easy to verify that Eq. (\ref{eq:F2TwistFieldSemiLocalZ2}) satisfies
the axioms (\ref{eq:FFAxiom1SemiLocalZ2}), (\ref{eq:FFAxiom2SemiLocalZ2})
and (\ref{eq:FFAxiom3SemiLocalZ2}). Analogously to Eq. \eqref{FFtil},
we have for $\tilde{\mathcal{T}}^{D}$ 
\begin{equation}
F_{2}^{\mathcal{T}^{D}|j,k}(\vartheta,n)=F_{2}^{\tilde{\mathcal{T}}^{D}|n-j,n-k}(\vartheta,n)\,.
\end{equation}

\section{$\mathbb{Z}_{2}$ branch-point twist field in the Ising model\label{sec:Ising}}

This section is devoted to the composite twist field of the Ising model. Clearly, the results for the FFs are interesting in their own
right, but the Ising model provides also several opportunities to test our results and some parts of the arguments on which our derivation
of the bootstrap equation relies. In particular, we can argue for the choice for the locality index $e^{i2\pi\gamma}=-1$ in the bootstrap
equations and we can demonstrate the existence of the spatial restriction of the $\mathbb{Z}_{2}$ symmetry. To do so, we borrow ideas from
\cite{gs-18} and use the lattice version of the Ising field theory with the Hamiltonian 
\begin{equation}
H=-J\sum_{i}\left(\sigma_{i}^{z}\sigma_{i+1}^{z}+h\sigma_{i}^{x}\right)\,,\label{HI}
\end{equation}
where $\sigma^{x/z}_i$ are the Pauli matrices. The conserved charge corresponding
to the $\mathbb{Z}_{2}$ symmetry is the fermion number parity $\hat{P}_{Q}$.
Here $\hat{Q}=\hat{Q}_{A}+\hat{Q}_{\bar{A}}$ is the fermion number
operator, which is clearly additive, and $\bar{A}$ denotes the
complement of the region $A$. Crucially, the parity operator has
eigenvalues $0$ or $1$ and the spacial restriction of this operator
is also additive in a $\text{mod }2$ sense, i.e., 
\begin{equation}
\hat{P}_{A}+\hat{P}_{\bar{A}}=\hat{P}\text{ mod }2\,,
\end{equation}
where we introduced the shorthand $\hat{P}_{Q_{A}}$ as $\hat{P}_{A}$.

An important quantity directly related to $\hat{P}$ is $(-1)^{\hat{Q}}$.
This quantity can be expressed in several ways allowing for the computation
of the symmetry resolved entropies in the critical point of the Ising
model \cite{gs-18} and in its off-critical, lattice version \cite{fg-20},
serving as valuable benchmark for our approach. Writing $\hat{P}$ as 
\begin{equation}
(-1)^{\hat{Q}_{A}}=\prod_{i\in A}\sigma_{i}^{x}\,,
\end{equation}
and introducing the disorder operators $\mu_{i}^{z}=\prod_{i\leq j}\sigma_{j}^{x}$
and $\mu_{i}^{x}=\sigma_{i}^{z}\sigma_{i+1}^{z}$ (satisfying the
same algebra of the Pauli matrices), we have 
\begin{equation}
(-1)^{\hat{Q}_{A}}=\prod_{i\in A}\sigma_{i}^{x}=\mu_{1}\mu_{\ell},\label{eq:Mu1MuL}
\end{equation}
when the region $A$ is a single interval from site $1$ to $\ell$. 
We recall that the disorder operator exists in the continuum limit as well. 
From Eq. (\ref{eq:Mu1MuL}) it is easy to deduce that the $\mathbb{Z}_{2}$
branch-point twist field must be related to fusion of the usual branch-point twist field and the disorder operator. 
This picture is confirmed explicitly at the critical point of the Ising field theory \cite{gs-18},
which corresponds to a conformal theory with central charge $c=\frac{1}{2}$.
The scaling dimension of $\mu$ is $\Delta_{\mu}=\bar{\Delta}_{\mu}=\frac{1}{16}$
and the symmetry resolved Rényi entropies for and interval of length
$\ell$ read \cite{gs-18} 
\begin{equation}
S_{n}(P_{A})=\ell^{-(n-1/n)/12}\frac{1}{2}\left(1+(-1)^{P_{A}}\ell^{-1/(4n)}\right)+\ldots\,,\label{eq:Z2EntropyCFTIsing}
\end{equation}
where $P_{A}$ is either $0$ or $1$. 
The disorder field $\mu$ has the property of changing boundary conditions from periodic to anti-periodic and vice versa.
This property corresponds to the locality index $e^{i2\pi\gamma}=-1$ in the residue and cyclic permutation axioms of
the bootstrap equations for its form factors in the massive theory. 
The value of this index confirms more rigorously that, for the Ising QFT, the $\mathbb{Z}_{2}$ branch-point
twist field form factors are obtained from Eqs. (\ref{eq:FFAxiom1}),
(\ref{eq:FFAxiom2}) and (\ref{eq:FFAxiom3}) with the insertion of
$e^{i2\pi\gamma}=-1$, resulting in Eqs. (\ref{eq:FFAxiom1SemiLocalZ2}),
(\ref{eq:FFAxiom2SemiLocalZ2}) and (\ref{eq:FFAxiom3SemiLocalZ2}).
We recall that the bootstrap equations have physically meaningful
solutions only for odd $n$ when 
\begin{equation}
\text{Tr}\left(\rho_{A}^{n}(-1)^{\hat{Q}_{A}}\right)=\text{Tr}\left(\rho_{A}^{n}(-1)^{n\hat{Q}_{A}}\right)\,,\label{eq:IsingTracewithQ}
\end{equation}
i.e. when the flux can be inserted on each of the $n$ copies.

The solutions for the bootstrap equations (\ref{eq:FFAxiom1SemiLocalZ2}),
(\ref{eq:FFAxiom2SemiLocalZ2}) and (\ref{eq:FFAxiom3SemiLocalZ2})
with locality index $e^{i2\pi\gamma}=-1$ for the $\mathbb{Z}_{2}$
branch-point twist field in the Ising model are easy to obtain. For
the minimal form factor we have 
\begin{equation}
F_{\text{min}}^{\mathcal{T}^{D}|1,1}(\vartheta,n)=-i\sinh\frac{\vartheta}{n},\label{fminI}
\end{equation}
from which $F_{2}^{\mathcal{T}^{D}|j,k}$ is obtained by (\ref{eq:F2TwistFieldSemiLocalZ2}).
As anticipated, and also confirmed later on in this section, the $\mathbb{Z}_{2}$
branch-point twist field can be regarded as a fusion of the conventional
twist field and the Ising disorder operator (on the same lines of
the composite fields for non-unitary theories \cite{bcd-15}). In
the off-critical theory, the FFs of both fields are non-vanishing
only for even particle numbers. It is therefore natural to expect
that $F_{k}^{\mathcal{T}^{D}}$ is also vanishing for odd $k$. Nevertheless,
even with the presence of FFs for odd particle numbers, their knowledge
would be not necessary for any of the considerations of this paper
\cite{cd-09} and, in fact, the VEV and the two-particle FFs encode
all the physics we are currently interested in.

The FFs for even particle number $F_{2k}^{\mathcal{T}^{D}}$ with
$2k\geq4$ can be written as a Pfaffian of the two-particle FF, similarly
to the case of the conventional branch-point twist field. For example,
considering the bootstrap equations for particle numbers $2k=4$ and
$6$, it can be directly verified that $F_{k}^{\mathcal{T}^{D}}$
indeed admits a Pfaffian representation. In particular, for $j_{1}\geq j_{2}\geq...\geq j_{2k}$,
one has 
\begin{equation}
F_{2k\text{ Ising}}^{\mathcal{T}^{D}|j_{1},...j_{2k}}(\vartheta_{1},...,\vartheta_{2k},n)=\langle\mathcal{T}_{n}^{D}\rangle\text{Pf}(W)\,,\label{eq:Pfaffain}
\end{equation}
where $W$ is a $2k\times2k$ anti-symmetric matrix with entries 
\begin{equation}
W_{lm}=\begin{cases}
\frac{F_{2}^{\mathcal{T}^{D}|j_{l},j_{m}}(\vartheta_{l}-\vartheta_{m},n)}{\langle\mathcal{T}_{n}^{D}\rangle} & m>l,\\
\left(-1\right)^{\delta_{j_{l},j_{m}}+1}\frac{F_{2}^{\mathcal{T}^{D}|j_{l},j_{m}}(\vartheta_{l}-\vartheta_{m},n)}{\langle\mathcal{T}_{n}^{D}\rangle} & m<l\,.
\end{cases}
\end{equation}
For general $k$, the Pfaffian structure \eqref{eq:Pfaffain} can
be shown by induction, following exactly the same lines of the proof
for conventional twist-fields \cite{cd-09b}. If the ordering of the
indices $j_{i}$ is not the canonical one, using the exchange axiom
\eqref{eq:FFAxiom1SemiLocalZ2} one can reshuffle the particles and
their rapidities to have $j_{1}\geq j_{2}\geq...\geq j_{2k}$ so to
apply \eqref{eq:Pfaffain}. When the order of particles with the same
replica index is left unchanged, the reshuffling does not introduce
any $\pm1$ factors.

Non-trivial checks of the solutions are provided by the limit  for $n\rightarrow1$ and the $\Delta$-theorem \cite{delta_theorem}. 
For $n\rightarrow1$, one expects to recover the form factors of the disorder operator; in particular for the two-particle case we expect
\begin{equation}
F_{2}^{D}(\vartheta)=i\langle\mu_{\text{Ising}}\rangle\tanh\frac{\vartheta}{2}\,,\label{eq:F2DisorderIsing}
\end{equation}
with $\langle\mu_{\text{Ising}}\rangle$ denoting the vacuum expectation value of $\mu_{\text{Ising}}$. 
The limit of the $\mathbb{Z}_{2}$ branch-point
twist field in the Ising model is 
\begin{equation}
\ \begin{split}\lim_{j,k,n\rightarrow1}F_{2}^{\mathcal{T}^{D}|j,k}(\vartheta,n)%
= & \lim_{j,k,n\rightarrow1}\frac{\langle\mathcal{T}_{n}^{D}\rangle\sin\frac{\pi}{n}}{2n\sinh\left(\frac{i\pi\left(2(j-k)-1\right)+\vartheta}{2n}\right)\sinh\left(\frac{i\pi\left(2(k-j)-1\right)-\vartheta}{2n}\right)}\frac{F_{\text{min}}^{\mathcal{T}^{D}|j,k}(\vartheta,n)}{F_{\text{min}}^{\mathcal{T}^{D}|1,1}(i\pi,n)}\\
= & -\langle\mathcal{T}_{1}^{D}\rangle\frac{-i\sinh\vartheta}{-\left(1+\cosh\vartheta\right)}\times\lim_{n\rightarrow1}\frac{\sin\frac{\pi}{n}}{-i\sinh\left(\frac{i\pi}{n}\right)}\\
= & \langle\mathcal{T}_{1}^{D}\rangle\frac{i\sinh\vartheta}{1+\cosh\vartheta}=i\langle\mathcal{T}_{1}^{D}\rangle\tanh\frac{\vartheta}{2}\,,
\end{split}
\label{FmuT}
\end{equation}
which equals (\ref{eq:F2DisorderIsing}) since $\langle\mu_{\text{Ising}}\rangle=\langle\mathcal{T}_{1}^{D}\rangle$
as shown in Appendix \ref{sec:Appendix-A-VEV}, where $\langle\mathcal{T}_{n}^{D}\rangle$
is determined too. Since also the FFs of the Ising disorder operator
can be cast in a Pfaffian form relying on the two-particle FF, the
match between the two-particle FFs implies that 
\begin{equation}
\lim_{\{j_{i}\},n\rightarrow1}F_{2k}^{\mathcal{T}^{D}|j_{1},...,j_{2k}}(\vartheta_{1},...,\vartheta_{2k},n)=F_{2k}^{\mu}(\vartheta_{1},...,\vartheta_{2k}).
\end{equation}

The second test for the validity of the solution is given by the
$\Delta$-theorem sum rule \cite{delta_theorem}. The $\Delta$-theorem
states that if at some length scale $R$ the theory can be described
by a CFT, then the difference of the conformal weight of an operator
$O$ and its conformal weight in the infrared (IR) limit can be calculated
as (if the integral converges) 
\begin{equation}
D(R)-\Delta^{IR}=-\frac{1}{4\pi\langle O\rangle}\int_{x^{2}>R}\mathrm{d^{2}}x\langle\Theta(x)O(0)\rangle_{c},\label{eq:DeltaTheoreM1}
\end{equation}
where $\Theta$ is the trace of the stress-energy tensor. 
Writing the spectral representation of (\ref{eq:DeltaTheoreM1}) in terms
of form factors, we have  \begin{equation}
D(r)-\Delta^{IR}=-\frac{1}{2\left\langle O\right\rangle }\sum_{n=1}^{\infty}\int\frac{\mathrm{d}\vartheta_{1}...\mathrm{d}\vartheta_{n}}{(2\pi)^{n}n!}\frac{e^{-rE_{n}}(1+E_{n}r)}{m^2E_{n}^{2}}F^{\Theta}\left(\vartheta_{1},\dots,\vartheta_{n}\right)F^{O}\left(\vartheta_{n},\dots,\vartheta_{1}\right)\:,\label{eq:DeltaTheoreM2}
\end{equation}
where $m$ is a mass scale $r=Rm$ and $mE_{n}$ are the $n$-particle
energies. For the case of the massive Ising model, the conformal weights
in the IR limit are zero. Hence taking $r=0$ in (\ref{eq:DeltaTheoreM2})
gives the UV conformal dimension of the operator $O$ as 
\begin{equation}
\Delta^{UV}=-\frac{1}{2\left\langle O\right\rangle }\sum_{k=1}^{\infty}\int\frac{\mathrm{d}\vartheta_{1}...\mathrm{d}\vartheta_{k}}{(2\pi)^{k}k!}E_{k}^{-2}m^{-2}F_{k}^{\Theta}\left(\vartheta_{1},\dots,\vartheta_{k}\right)F_{k}^{O}\left(\vartheta_{k},\dots,\vartheta_{1}\right)\,.\label{eq:DeltaTheoreM3}
\end{equation}
In the Ising field theory, as well as in its $n$-copy version, the field $\Theta$ has non-vanishing form factors only in the two-particle
sector, so the sum is terminated by the $k=2$ contribution. 
After easy manipulations, the same as in Ref. \cite{ccd-08} for the conventional twist fields, Eq. (\ref{eq:DeltaTheoreM3}) 
for the $\mathbb{Z}_{2}$ branch-point twist field can be written  as 
\begin{equation}
\Delta^{\mathcal{T}_{n}^{D}}=-\frac{n}{32\pi^{2}m^{2}\left\langle \mathcal{T}_{n}^{D}\right\rangle }\int\mathrm{d}\vartheta\frac{F_{2}^{\Theta|1,1}\left(\vartheta\right)F_{2}^{\mathcal{T}^{D}|1,1}(\vartheta,n)^{*}}{\cosh^{2}\left(\vartheta/2\right)}\,,\label{eq:DeltaTheoreZ2TwistFieldIsing}
\end{equation}
with 
\begin{equation}
F_{2}^{\Theta|1,1}\left(\vartheta\right)=-2\pi im^{2}\sinh\frac{\vartheta}{2}\,.
\end{equation}
We evaluated the integral in (\ref{eq:DeltaTheoreZ2TwistFieldIsing}) numerically for many integer odd $n$ using the FF \eqref{eq:F2TwistFieldSemiLocalZ2}. 
We found that the numerical calculated integrals match perfectly the prediction $\frac{c}{24}\left(n-n^{-1}\right)+\frac{\Delta}{n}$
\cite{gs-18} with $c=\frac{1}{2}$ and $\Delta=\frac{1}{16}$ for
all the considered $n$. Such perfect agreement is a strong evidence
for the correcteness of the FF $F_{2}^{\mathcal{T}^{D}|1,1}(\vartheta,n)$ in Eq. \eqref{eq:F2TwistFieldSemiLocalZ2}.
%The numerical data can be found in the following table:
%\begin{center}
%\begin{tabular}{|c|c|c|}
%\hline 
%n & $\frac{c}{24}\left(n-n^{-1}\right)+\frac{\Delta}{n}$ & $\Delta$-theorem\tabularnewline
%\hline 
%\hline 
%1 & 1/16 & 1/16\tabularnewline
%\hline 
%3 & 11/144 & 11/144\tabularnewline
%\hline 
%5 & 9/80 & 9/80\tabularnewline
%\hline 
%7 & 17/112 & 17/112\tabularnewline
%\hline 
%\end{tabular}
%\par\end{center}

\section{$\mathbb{Z}_{2}$ branch-point twist field in the sinh-Gordon model\label{sec:ShG} }

As shown in section \ref{sec:Z2TwistFields}, the solution of the
bootstrap equations (\ref{eq:FFAxiom1SemiLocalZ2}), (\ref{eq:FFAxiom2SemiLocalZ2})
and (\ref{eq:FFAxiom3SemiLocalZ2}) is also possible for the ShG model.
These equations include the locality factor $e^{i2\pi\gamma}=-1$
and their solution differs from the FFs of the conventional twist
fields by an additional CDD factor (\ref{eq:fCDDSemiLocalZ2}) and
a different sign prescription in (\ref{eq:FDMinBasicProp}). As seen
in the previous section, the corresponding solution for the Ising
model can be associated with the $\mathbb{Z}_{2}$ symmetry resolution
of entropies. Nevertheless, the question of whether the symmetry resolution
is possible, i.e., some/any reduced density matrices commute with
the operator corresponding to the $\mathbb{Z}_{2}$ symmetry is a
rather difficult one for the ShG model. In the following, we present
a series of arguments to claim that such a symmetry resolution is
plausible at least for a single interval in the ground state of the
model.

The first argument is based on the application of the Bisognano-Wichmann
theorem \cite{BisognanoWichmann} to the ShG model. This theorem states
that for the ground state of a spatially infinite relativistic QFT,
the reduced density matrix of a half-infinite line can be written
as 
\begin{equation}
\rho\propto\exp(-2\pi K),
\end{equation}
with the modular (or entanglement) Hamiltonian $K$ 
\begin{equation}
K=\int_{0}^{\infty}\text{d}x\,x\mathcal{H}_{\text{}}[\varphi(x)]\,,
\end{equation}
where $\mathcal{H}_{\text{}}$ is the hamiltonian density. For the
ShG model, the hamiltonian density $\mathcal{H}_{\text{ShG}}$ is
invariant under the $\mathbb{Z}_{2}$ transformation $\varphi\rightarrow-\varphi$,
hence $K$ and $\rho$ commute with the $\mathbb{Z}_{2}$ symmetry
operation. The ShG model is a massive theory, and hence it is plausible
that the RDM of an interval still commutes with the symmetry
operation, at least for long enough distance, which is the case for
which we eventually apply the novel form factors.

A second argument is given by the conformal limit of the ShG model,
which is a free massless conformal boson. For the ground state of
CFTs, the modular Hamiltonian is also known for a single interval
of length $2R$ \cite{CFTInterval1,CFTInterval2,ct-16} and reads
\begin{equation}
K=\int_{-R}^{R}\text{d}x\,\frac{R^{2}-x^{2}}{2R}\mathcal{H}_{\text{CFT}}[\varphi(x)]\,.
\end{equation}
The Hamiltonian density of the free massless boson is again invariant
under the $\mathbb{Z}_{2}$ transformation $\varphi\rightarrow-\varphi$,
and, repeating the previous reasoning, the possibility of the symmetry
resolution is justified in the UV regime.

Finally, we consider another limit of the ShG theory, namely when
$B=1+i\frac{2}{\pi}\Theta_{0}$ with $\Theta_{0}\rightarrow\infty$.
As already noted, in this limit the form factors of the ShG model
reduce to those of the Ising model. As shown below, $\text{\ensuremath{F_{2,\text{ShG}}^{\mathcal{T}^{D}|j,k}}(\ensuremath{\vartheta},n)}$
is no exception to this rule, because the CDD factor
$f_{\text{CDD}}(\vartheta)$ is the same for the Ising and ShG models
and
\begin{equation}
F_{2,\text{ShG}}^{\mathcal{T}|j,k}(\vartheta,n)\rightarrow F_{2, {\rm Ising}}^{\mathcal{T}|j,k}(\vartheta,n)\,.
\end{equation}
 Consequently, the limit 
\begin{equation}
F_{2,\text{ShG}}^{\mathcal{T}^{D}|j,k}(\vartheta,n)\rightarrow F_{2, {\rm Ising}}^{\mathcal{T}^{D}|j,k}(\vartheta,n)\label{eq:Z2TwistFieldRoamingCollapse}
\end{equation}
holds: this link between the two models provides another evidence for the plausibility of a $\mathbb{Z}_{2}$ symmetry resolution of
the ShG model.

It is now worth studying some features of these FFs and in particular
the two-particle one, $F_{2,\text{ShG}}^{\mathcal{T}^{D}|j,k}(\vartheta,n)$.
First of all, similarly to the Ising model, it is expected that $F_{k,\text{ShG}}^{\mathcal{T}^{D}}$
vanishes for odd $k$. The reason is always the same: the $\mathbb{Z}_{2}$
branch-point twist field can be regarded as a fusion of the conventional
ShG twist field and the ShG disorder operator or twist field (which
should not be mistaken for the branch-point twist field). In the off-critical
theory, the FFs of both fields are non-vanishing only for even particle
numbers. %It is, however, stressed again that for the considerations of this paper the FFs for odd particle numbers are not important \cite{cd-09}.
Considering now the two-particle FF solution, an interesting insight
is given by the $n\rightarrow1$ limit of $F_{2,\text{ShG}}^{\mathcal{T}^{D}|j,k}(\vartheta,n)$.
The first few form factors of the ShG twist field are known and were
constructed in \cite{Roaming FF}. This field can be identified with
the off-critical version of the twist field of the massless free boson
theory, where a unique field exists which changes the boundary condition
of the boson field from periodic to anti-periodic and vice versa.
This field has conformal weight $\Delta=1/16=0.0625$ \cite{AppCFT}
and can be regarded as bosonic analogue of the fermionic disorder
operator.

We now show that in the limit $n\rightarrow1$, $F_{2,\text{ShG}}^{\mathcal{T}^{D}|j,k}(\vartheta,n)$
coincides with $F_{2,\text{ShG}}^{D}(\vartheta)$, where $F_{2,\text{ShG}}^{D}(\vartheta)$
is the two-particle form factor of ShG twist field (again, the disorder
operator, not the branch-point one). According to Ref. \cite{Roaming FF},
\begin{equation}
F_{2,\text{ShG}}^{D}(\vartheta_{1},\vartheta_{2})=-2\langle\mu_{\text{ShG}}^{D}\rangle\frac{\sqrt{e^{\vartheta_{1}+\vartheta_{2}}}}{e^{\vartheta_{1}}+e^{\vartheta_{2}}}f_{11,\text{ShG}}(\vartheta_{1}-\vartheta_{2})\,,\label{eq:2PTShGTwistField}
\end{equation}
where $f_{11,\text{ShG}}$ is defined in Eq. (\ref{eq:f11}), $\langle\mu_{\text{ShG}}^{D}\rangle$
is the vacuum expectation value of the ShG twist field, and though
not manifest from its form, (\ref{eq:2PTShGTwistField}) depends only
on the difference of $\vartheta_{1}$ and $\vartheta_{2}$.
From $F_{2,\text{ShG}}^{\mathcal{T}^{D}|j,k}$ we can proceed as 
\begin{equation}
\begin{split}\lim_{j,k,n\rightarrow1} & F_{2,\text{ShG}}^{\mathcal{T}^{D}|j,k}(\vartheta,n)=\\
= & \lim_{j,k,n\rightarrow1}\frac{\langle\mathcal{T}_{n,\text{ShG}}^{D}\rangle\sin\frac{\pi}{n}}{2n\sinh\left(\frac{i\pi\left(2(j-k)-1\right)+\vartheta}{2n}\right)\sinh\left(\frac{i\pi\left(2(k-j)-1\right)-\vartheta}{2n}\right)}\frac{\cosh\left(\frac{\vartheta}{2n}\right)F_{\text{min,ShG}}^{\mathcal{T}|j,k}(\vartheta,n)}{\cosh\left(\frac{i\pi}{2n}\right)F_{\text{min,ShG}}^{\mathcal{T}|1,1}(i\pi,n)}\\
= & -\langle\mathcal{T}_{1,\text{ShG}}^{D}\rangle\frac{\cosh\left(\frac{\vartheta}{2}\right)F_{\text{min}}^{\mathcal{T}|j,k}(\vartheta,1)}{\left(1+\cosh(\vartheta)\right)F_{\text{min,ShG}}^{\mathcal{T}|1,1}(i\pi,1)}\times\lim_{n\rightarrow1}\frac{\sin\frac{\pi}{n}}{\cosh\left(\frac{i\pi}{2n}\right)}\\
= & -2\langle\mathcal{T}_{1,\text{ShG}}^{D}\rangle\frac{\cosh\left(\frac{\vartheta}{2}\right)F_{\text{min},ShG}^{\mathcal{T}|j,k}(\vartheta,1)}{\left(1+\cosh(\vartheta)\right)F_{\text{min,ShG}}^{\mathcal{T}|1,1}(i\pi,1)}=-2\langle\mathcal{T}_{1,\text{ShG}}^{D}\rangle\frac{\cosh\left(\frac{\vartheta}{2}\right)}{\left(1+\cosh(\vartheta)\right)}f_{11,\text{ShG}}(\vartheta)\,.
\end{split}
\end{equation}
%$\frac{F_{\text{min},\text{ShG}}^{\mathcal{T}|j,k}(\vartheta_{1}-\vartheta_{2},1)}{F_{\text{min,ShG}}^{\mathcal{T}|1,1}(i\pi,1)}$
%equals $f_{11,\text{ShG}}(\vartheta_{1}-\vartheta_{2})$ and anticipating
At this point, we should just use $\langle\mathcal{T}_{1,\text{ShG}}^{D}\rangle=\langle\mu_{\text{ShG}}^{D}\rangle$
and $\frac{\sqrt{e^{\vartheta_{1}+\vartheta_{2}}}}{e^{\vartheta_{1}}+e^{\vartheta_{2}}}=\frac{\cosh\left(\frac{\vartheta_{1}-\vartheta_{2}}{2}\right)}{1+\cosh(\vartheta_{1}-\vartheta_{2})}$
to prove our claim.

{ 
\begin{table}[t]
\begin{centering}
\subfloat[{$B=0.4$}]{%
\begin{centering}
\begin{tabular}{|c|c|c|c|}
\hline 
$n$  & $\frac{c}{24}\left(n-n^{-1}\right)+\frac{\Delta}{n}$  & $\frac{c}{24}\left(n-n^{-1}\right)$  & two-particle contribution\tabularnewline
\hline 
\hline 
1  & 0.0625  & 0  & 0.0664945\tabularnewline
\hline 
3  & 0.131944  & 0.111111  & 0.137754\tabularnewline
\hline 
5  & 0.2125  & 0.2  & 0.221387 \tabularnewline
\hline 
7  & 0.294643  & 0.285714  & 0.306779 \tabularnewline
\hline 
\end{tabular}
\par\end{centering}
{\small{}}{\small \par}}
\par\end{centering}{\small \par}
\begin{centering}
{\small{}}\subfloat[$B=0.6$]{%
\begin{centering}
\begin{tabular}{|c|c|c|c|}
\hline 
$n$  & $\frac{c}{24}\left(n-n^{-1}\right)+\frac{\Delta}{n}$  & $\frac{c}{24}\left(n-n^{-1}\right)$  & two-particle contribution\tabularnewline
\hline 
\hline 
1  & 0.0625  & 0  & 0.0674768\tabularnewline
\hline 
3  & 0.131944  & 0.111111  & 0.138998\tabularnewline
\hline 
5  & 0.2125  & 0.2  & 0.223242\tabularnewline
\hline 
7  & 0.294643  & 0.285714  & 0.309292 \tabularnewline
\hline 
\end{tabular}
\par\end{centering}
{\small{}}{\small \par}}
\par\end{centering}{\small \par}
{\small{}\caption{\label{tab:--paired-state}The two-particle contributions of the $\Delta$-theorem
sum rule compared with the expected conformal dimension of $\mathbb{Z}_{2}$
and conventional branch-point twist fields in ShG model.}
\label{tab1}
}{\small \par}
\end{table}
{\small \par}

Based on this finding, it is natural to expect that the UV scaling
dimension of the ShG $\mathbb{Z}_{2}$ twist field is $\frac{c}{12}\left(n-n^{-1}\right)+\frac{\Delta}{n}$
with $c=1$ and $\Delta=1/16$. We close this section showing that
the $\Delta$-theorem \cite{delta_theorem} is consistent with this
assumption. Unlike for the Ising model, the form factors of the stress
energy tensor in the ShG model are non-vanishing for the $k=4,6,...$-particle
sectors. In the integral formula of the $\Delta$-theorem only the
two-particle contribution is included and so it is not expected to
be exact, but still to be a very good approximation. We calculated
numerically such total 2-particle contribution for several $B$ confirming
such expectation. In the table \ref{tab1}  we show such comparison for $B=0.4$
and $0.6$.
Notice that the two-particle contribution is always slightly larger
than the expected total value and the difference
is larger for larger $B$ (up to $B=1$), which is a general feature
of the ShG model. This is very similar to what observed for the conventional
twist field in Ref. \cite{ccd-08} and also the difference is of the
same order of magnitude. We stress that the fact that the offset is
positive is an error (as the non-ideal name `sum rule' would
suggest): in Eq. \eqref{eq:DeltaTheoreZ2TwistFieldIsing} we do not
have the integral of a positive defined quantity.

\section{General results on $\mathbb{Z}_{2}$ symmetry resolved entropy in massive QFT}
\label{sec:gen}

In this section, we first present some basic and elementary facts about
the symmetry resolved entanglement entropies for an arbitrary theory
with $\mathbb{Z}_{2}$ symmetry and then exploit the QFT scaling form
to derive some general results valid for arbitrary massive QFTs. For
conciseness in writing formulas, in this and in the following section,
we switch to the notation $+$ and $-$ for the quantum numbers that
replace $0$ and $1$ respectively: since we focus on ${\mathbb{Z}}_{2}$
symmetry there is no ambiguity with this notation. Let us recall the
definition of the symmetry resolved partition functions (\ref{eq:CalZn})
in terms the charged moments (\ref{eq:Zn}): 
\begin{equation}
\mathcal{Z}_{n}(\pm)=\frac{1}{2}\left(Z_{n}(0)\pm Z_{n}(1)\right),%,
\end{equation}
where 
\begin{equation}
Z_{n}(0)=\text{Tr}\rho_{A}^{n}\,,
\end{equation}
and 
\begin{equation}
Z_{n}(1)=\text{Tr}\left[\rho_{A}^{n}\exp\left(i\pi\hat{P}_{A}\right)\right]\,. %=\text{Tr}\rho_{A}^{n}(-1)^{\hat{Q}_{A}}\,.
\end{equation}
Here $Z_{n}(1)$ is the charged moment associated with the two-point
function of $\mathbb{Z}_{2}$ twist field. From Eq. \eqref{eq:vNE},
the symmetry resolved Rényi entropies can be written as (recall that
$Z_{1}(0)=1$ by normalisation) 
\begin{equation}
S_{n}(\pm)=\frac{1}{1-n}\ln\left[\frac{{\cal Z}_{n}(\pm)}{{\cal Z}_{1}^{n}(\pm)}\right]=\frac{1}{1-n}\ln\left[\frac{Z_{n}(0)\pm Z_{n}(1)}{\left(1\pm Z_{1}(1)\right)^{n}}2^{n-1}\right].%,
\label{eq:S0S1Renyi}
\end{equation}
In any 2D QFT, the two (charged and neutral) moments entering in the
Rényi entropies of an interval $A=[u,v]$ (with $\ell=v-u$) are written
as 
\begin{eqnarray}
Z_{n}(0) & = & \text{Tr}\rho_{A}^{n}=\zeta_{n}\varepsilon^{2d_{n}}\langle\mathcal{T}_{n}(u,0)\tilde{\mathcal{T}}_{n}(v,0)\rangle\,,\label{eq:Zn0TwoPt}\\
Z_{n}(1) & = & \text{Tr}[\rho_{A}^{n}(-1)^{n\hat{Q}_{A}}]=\zeta_{n}^{D}\varepsilon^{2d_{n}^{D}}\langle\mathcal{T}_{n}^{D}(u,0)\tilde{\mathcal{T}}_{n}^{D}(v,0)\rangle\,,\label{eq:Zn1TwoPt}
\end{eqnarray}
where $\varepsilon$ is the UV regulator, $\zeta_{n}^{D}$ and $\zeta^{D}$
the normalisation constants of the composite and conventional twist
fields, respectively, and $d_{n}$ and $d_{n}^{D}$ their dimensions,
given as 
\begin{equation}
d_{n}=%\Delta^{\mathcal{T}_{n}}+\bar{\Delta}^{\mathcal{T}_{n,}}=\Delta^{\mathcal{\tilde{T}}_{n}}+\bar{\Delta}^{\tilde{\mathcal{T}}_{n}}=
2\Delta^{\mathcal{T}_{n}}=\frac{c}{12}\left(n-n^{-1}\right),\qquad d_{n}^{D}=%\Delta^{\mathcal{T}_{n}^{D}}+\bar{\Delta}^{\mathcal{T}_{n}^{D}}=\Delta^{\mathcal{\tilde{T}}_{n}^{D}}+\bar{\Delta}^{\tilde{\mathcal{T}}_{n}^{D}}=
2\Delta^{\mathcal{T}_{n}^{D}}=2\Delta^{\mathcal{T}_{n}}+2\frac{\Delta}{n}=\frac{c}{12}\left(n-n^{-1}\right)+2\frac{\Delta}{n}\,,
\end{equation}
where $\Delta$ is the dimension of the field that fuses with the
conventional twist-field to give the ${\mathbb{Z}}_{2}$ composite
one (e.g. the disorder operator in the Ising model or ShG with dimension
$\Delta=1/16$).

It is then clear that in the two symmetry resolved entropies \eqref{eq:S0S1Renyi},
in the QFT regime $\varepsilon\ll1$, we have $Z_{n}(1)\ll Z_{n}(0)$
because $\Delta$ is positive. Hence we find the `trivial', yet general,
result 
\begin{equation}
S_{n}(\pm)=S_{n}-\ln2+\mathcal{O}(\varepsilon^{\frac{4\Delta}{n}}),%,
\label{finalEQ}
\end{equation}
where $S_{n}$ is the total R\'enyi entropy. % (the factor $\ln \epsilon$ in the correction in \eqref{finalEQ} is inherited from the leading term $S_n$). 
For general $n$ the total Rényi entropy is known for some models,
see e.g. \cite{ccd-08,cd-09}, but its form is rather cumbersome.
Instead, in the von Neumann limit, the result considerably simplifies
in a generic massive model to \cite{ccd-08} 
\begin{equation}
S=-\frac{c}{3}\ln m\varepsilon+U-\frac{1}{8}K_{0}(2m\ell)+\cdots\,,
\end{equation}
where $U$ is a model dependent constant (e.g. calculated for the
Ising model in \cite{ccd-08}) and $m$ the mass of the lightest particle
of the field theory. We anticipate that for $n=1$, the corrections
in \eqref{finalEQ} gets multiplied by $\ln\varepsilon$, as we shall
see later in this section.

In spite of its triviality, Eq. \eqref{finalEQ} shows that in a general
${\mathbb{Z}}_{2}$-symmetric QFT there is equipartition of entanglement
at the leading order in $\varepsilon$. The term $-\ln2$ which sums
to the total entropy is a consequence of the fluctuation entropy in
Eq. (\ref{eq:SfSc}). Indeed, for $\varepsilon\to0$, we have $p(0)=\mathcal{Z}_{1}(0)=p(1)=\mathcal{Z}_{1}(1)=\frac{1}{2}$,
and hence the number entropy is just $S^{f}=-\frac{2}{2}\ln\frac{1}{2}$.
Consequently, in Eq. (\ref{eq:SfSc}) we have 
\begin{equation}
S=\frac{S(+)+S(-)}{2}-\frac{2}{2}\ln\frac{1}{2}=S\,.
\end{equation}

However, this is not the end of the story. Eq. \eqref{finalEQ} with
\eqref{eq:S0S1Renyi} shows that there are corrections to entanglement
equipartition that are calculable within the integrable QFT framework
of this paper. In fact, expanding Eq. \eqref{eq:S0S1Renyi} for $Z_{n}(1)\ll Z_{n}(0)$
we have 
\begin{equation}
S_{n}(\pm)=S_{n}-\ln2\pm\frac{1}{1-n}\left(\frac{Z_{n}(1)}{Z_{n}(0)}-nZ_{1}(1)\right)+\cdots.\label{eq:SRenyiApprox}
\end{equation}
%where, for conciseness, we replaced $0$ and $1$ with $+$ and $-$ respectively. 
Notice that for generic $n>1$, the ratio $\frac{Z_{n}(1)}{Z_{n}(0)}$
is proportional to $\varepsilon^{4\Delta/n}$ while $Z_{1}(1)\propto\varepsilon^{4\Delta}$
and so the former is the leading correction. The two corrections become
of the same order in the physically relevant limit $n\to1$. Notice
that these corrections are very much reminiscent of the unusual corrections
to the scaling \cite{ccen-09,cc-10} as calculated in massive theories
\cite{ccp-10}. This is not a coincidence since also unusual corrections
in field theory come from the fusion of the twist field with a relevant
operator \cite{cc-10}.

%The equipartition of entanglement is then broken by the subleading quantity
%\begin{equation}
%\delta S_+\equiv \frac 1{1-n}\left(\frac{Z_n(1)}{Z_n(0)}- n Z_1(1)\right).
%\end{equation} 
Exploiting Eqs. \eqref{eq:Zn0TwoPt} and \eqref{eq:Zn1TwoPt}, we
have 
\begin{equation}
\frac{Z_{n}(1)}{Z_{n}(0)}=\varepsilon^{4\Delta/n}\frac{\zeta_{n}^{D}}{\zeta_{n}}\frac{\langle\mathcal{T}_{n}^{D}(u,0)\tilde{\mathcal{T}}_{n}^{D}(v,0)\rangle}{\langle\mathcal{T}_{n}(u,0)\tilde{\mathcal{T}}_{n}(v,0)\rangle}.
\end{equation}
This expression provides the leading term breaking equipartition of entanglement for $n>1$. 
With the exception of the normalisation amplitudes $\zeta_n$ and $\zeta_n^D$ which depend on the precise UV regularisation of the theory (lattice in the following),
all the quantities entering in the above ratio are in principle accessible to the bootstrap approach and calculable
once the FFs are known. 

In the von Neumann limit, $n\to1$, it is convenient to write down
some general formula before taking the limit $Z_{n}(1)\ll Z_{n}(0)$.
In general we have 
\begin{equation}
%\begin{split}\end{split}
S(\pm)=-\frac{\partial}{\partial n}\left[\frac{Z_{n}(0)\pm Z_{n}(1)}{\left(1\pm Z_{1}(1)\right)^{n}}2^{n-1}\right]_{n=1}=\frac{S\pm s(1)}{1\pm Z_{1}(1)}+\ln(1\pm Z_{1}(1))-\ln{2},%S(1)=
%{split}
\label{eq:S0S1}
\end{equation}
where, once again, $S$ is the total entropy, and we defined 
\begin{equation}
s(1)\equiv-\lim_{n\rightarrow1}\frac{\partial}{\partial n}\text{Tr}\rho_{A}^{n}(-1)^{\hat{Q}_{A}}\,.
\end{equation}
We now take the limit $Z_{n}(1)\ll Z_{n}(0)$ (implying $Z_{1}(1)\ll1$
and $s(1)\ll S$), obtaining 
\begin{equation}
%\begin{split}\end{split}
S(\pm)=S-\ln2\mp SZ_{1}(1)\pm Z_{1}(1)\pm s(1)+o(\varepsilon^{4\Delta})\,.\label{eq:S0S1exp}
\end{equation}
Here the terms $SZ_{1}(1)$ and $s(1)$ behave as $\varepsilon^{4\Delta}\ln\varepsilon$,
while $Z_{1}(1)$ is proportional to $\varepsilon^{4\Delta}$. Hence
the breaking of equipartition of the von Neumann entanglement entropy
at leading order is fully encoded in the quantities $Z_{1}(1)$ and
$s(1)$ defined above. These are obtainable in the FF approach and
we will show with an explicit calculation for the Ising field theory
in the next section. Although these terms breaking equipartition are
vanishing in the field theory limit, they can be straightforwardly
evaluated in any numerical computation (e.g. taking the difference
$S(+)-S(-)$ which cancels the leading term and isolate the correction).
Such numerical computations can be verified against the predictions
after having identified (as e.g. done in the next section for the
Ising model) or fitted the non-universal UV cutoff $\varepsilon$.
The remaining difference is a universal scaling function of $m\ell$
which is calculable within the FF approach, as again shown for the
Ising model in the forthcoming section.

%As in any QFT calculation, it is necessary to introduce a 
%$\varepsilon$ to obtain finite quantities. Nevertheless, for large
%distances we have 
%\begin{equation}
%\lim_{\varepsilon\rightarrow0}\frac{Z_{n}(1)}{Z_{n}(0)}=0\,,\label{eq:vanishingZ1}
%\end{equation}
%which is valid for the von Neumann entropy too yielding the equipartition
%of entropy up to corrections $\varepsilon^{1/(4n)}$ for Rényi entropies
%and $\varepsilon^{1/4}$ for the von Neumann entropy. This exponent
%solely depends on the UV dimension of the $\mathbb{Z}_{2}$ twist
%field, hence the same result is expected to hold for the ShG model
%as well and, in fact, for the non-symmetry breaking ground state of
%any 1+1D QFT with a $\mathbb{Z}_{2}$ symmetry and with free boson/fermion
%UV limiting theories.
%It is easy to see that this finding, namely the equipartition of the
%entropy and the $\mathcal{O}(\varepsilon^{\frac{1}{4}}\ln\varepsilon)$
%type correction hold also for the ground state of ShG model. Corrections
%to the equipartition of the entropy depending on the UV regulator
%have been recently reported in the massive free complex boson and
%free Dirac fermion theories \cite{mdc-20b} for the U(1) continuous
%symmetry.

\section{Entropies from two-point functions of the $\mathbb{Z}_{2}$ branch-point
twist field in the Ising model\label{sec:CalculatingEntropies} }

In this section we show how the calculation of the symmetry resolved
von Neumann entropies can be carried out based on the knowledge of
the $\mathbb{Z}_{2}$ branch-point twist field. We restrict our analysis
to an interval in the ground state of Ising model in the paramagnetic
phase, where the entropies can be calculated from the two-point functions
of the conventional and composite twist fields. Our findings will
be checked against the continuum limit of the existing results for
the lattice model \cite{fg-20}. The calculation follows the logic
of Ref. \cite{ccd-08} including also steps like the determination
of the vacuum expectation value of the $\mathbb{Z}_{2}$ branch-point
twist-field, the analytic continuation of the charged moments, and
some further technical, but relatively straightforward, algebraic
manipulations. The interested reader is encouraged the consult to
corresponding appendices, where we report all the steps not strictly
necessary to follow the main ideas.

The symmetry resolved entropies for one interval can be calculated in terms of two-point function of the composite and conventional twist
fields just plugging \eqref{eq:Zn1TwoPt} and \eqref{eq:Zn0TwoPt} into \eqref{eq:S0S1Renyi} and \eqref{eq:S0S1} (or even to \eqref{eq:SRenyiApprox}
and \eqref{eq:S0S1exp}). 
The partition sum $Z_{n}(0)$, i.e., Eq. \eqref{eq:Zn0TwoPt}, determines the total entropy and all the required quantities for its calculation
$S_n$ were derived in Ref. \cite{ccd-08} (including the analytic continuation). 
Concerning $Z_n(1)$ in Eq. \eqref{eq:Zn0TwoPt}, the two-point function of the $\mathbb{Z}_{2}$ twist field and its vacuum
expectation value can be determined using purely QFT techniques, % as presented in Appendix \ref{sec:Appendix-A-VEV},
whereas the proportionality constant can be fixed by comparing the lattice and QFT results. 
Explicitly, we rewrite 
\begin{equation}
Z_{n}(1)=
\zeta_{n}^{D}(m\varepsilon)^{2d_{n}^{D}}[m^{-2d_{n}^{D}}\langle\mathcal{T}_{n}^{D}(u,0)\tilde{\mathcal{T}}_{n}^{D}(v,0)\rangle]
\equiv \zeta_{n}^{D}(m\varepsilon)^{2d_{n}^{D}}[ (m^{-2d_{n}^{D}}\langle\mathcal{T}_{n}^{D}\rangle^2)] H_n(m\ell) 
\,,\label{zn1}
\end{equation}
so that $m^{-2d_{n}^{D}}\langle\mathcal{T}_{n}^{D}(u,0)\tilde{\mathcal{T}}_{n}^{D}(v,0)\rangle$ is dimensionless and universal. 
Furthermore, we isolated the vacuum expectation value and defined the universal function $H_n(m\ell)$. 
Once again, we stress that both $\zeta_n^D$ and $(m^{-2d_{n}^{D}}\langle\mathcal{T}_{n}^{D}\rangle^2)$ are just numerical amplitudes, 
i.e. independent of $m$ and $\ell$. 

Focusing now on the von Neumann entropy, we only need to know Eqs. \eqref{eq:Zn0TwoPt} and \eqref{eq:Zn1TwoPt}
%, including the vacuum expectation value of the fields and the proportionality constants $\zeta_{n}^{D}$ and $\zeta_{n}$ 
in the vicinity of $n=1$. 
Hence, on top of $Z_1(1)$ given by Eq. \eqref{zn1}, we also need its derivative in $1$ which we rewrite as 
%To make more transparent the calculation and the list of needed quantities, let us rewrite explicitly the two
%quantities we primarily need to compute for the symmetry resolved
%von Neumann entropy in the Ising QFT, which, according to \eqref{eq:S0S1exp}, are 
%\begin{equation}
%Z_{1}(1)=%\lim_{n\rightarrow1}
%\zeta_{1}^{D}(m\varepsilon)^{2d_{1}^{D}}m^{-2d_{1}^{D}}\langle\mathcal{T}_{1}^{D}(u,0)\tilde{\mathcal{T}}_{1}^{D}(v,0)\rangle\,,\label{z11}
%\end{equation}
%and 
\begin{multline}
s(1)=-\lim_{n\rightarrow1}\frac{\partial}{\partial n}\left(\zeta_{n}^{D}(m\varepsilon)^{2d_{n}^{D}}m^{-2d_{n}^{D}}\langle\mathcal{T}_{n}^{D}(u,0)\tilde{\mathcal{T}}_{n}^{D}(v,0)\rangle\right)=\\
-Z_{1}(1)\lim_{n\to1}\left[\frac{\text{d}\ln\zeta_{n}^{D}}{\text{d}n}+2\frac{\text{d}d_{n}^{D}}{\text{d}n}\ln(m\varepsilon)+
\frac{\partial}{\partial n}\ln (m^{-2d_{n}^{D}}\langle\mathcal{T}_{n}^{D}\rangle^2) + \frac{\partial \ln H_n(m\ell) }{\partial n}\right]\,.\label{eq:s(1)}
\end{multline}
We stress that the entire $\ell$ dependence, which is the main focus of this approach, is fully encoded in the universal function $H_n(m\ell)$.  
The easiest part of the above expressions is $\frac{\text{d}d_{n}^{D}}{\text{d}n}$,
i.e. 
\begin{equation}
\lim_{n\to1}2\frac{\text{d}d_{n}^{D}}{\text{d}n}=-\frac{1}{12}\,.
\end{equation}
In the two following subsections we explicitly calculate all amplitudes
and two-point functions of composite twist fields.

\subsection{Computation of the amplitudes}
\label{sec:ampl}

In Eqs. \eqref{zn1} and \eqref{eq:s(1)}, a first ingredient yet
to be calculated is the amplitude $\zeta_{n}^{D}$. For $n=1$ there
is a straightforward way to get it, exploiting the fact that $\mathcal{T}_{1}^{D}$
equals the standard disorder operator. We can then write 
\begin{equation}
%\lim_{\text{QFT}}Z_{1}^{\text{Lat}}(1)=\lim_{\text{QFT}}\text{Tr}_{A}\rho_{A}(-1)^{\hat{Q}}=
\lim_{\text{QFT}}\langle\mu_{1}\mu_{j}\rangle_{{\rm Lat}}=\zeta_{1}^{D}\varepsilon^{2d_{1}^{D}}\langle\mathcal{T}_{1}^{D}(0,0)\tilde{\mathcal{T}}_{1}^{D}(aj,0)\rangle\,,\label{LitsOfLimits}
\end{equation}
where the expectation values $\langle\cdot\rangle_{{\rm Lat}}$ are
taken on the ground state of the lattice Hamiltonian \eqref{HI} with
lattice spacing $a$. We recall $d_{1}^{D}=\frac{1}{8}$. Here ${\displaystyle \lim_{\text{QFT}}}$
denotes the continuum limit of the lattice model, which is 
\begin{equation}
J\to\infty,\qquad a\to0,\qquad h\to1,
\end{equation}
with 
\begin{equation}
m=2J|h-1|,\qquad2Ja=v=1\,,\label{IsingQFTLimit}
\end{equation}
where $m$ is the field theoretical mass and $v$ the velocity of
light, that in our notation is $1$. The continuum limit $\mu(x)$
of the disorder operator ${\displaystyle \mu_{j}^{x}\equiv\prod_{j'=1}^{j}\sigma_{j'}^{x}}$
is \cite{MuVEV} 
\begin{equation}
\mu(ja)=\bar{s}J^{\frac{1}{8}}\mu_{j}^{x}\,,%{equation}
{\qquad{\rm with}}\qquad%!TEXencoding=UTF-8Unicode
\bar{s}=2^{\frac{1}{12}}e^{-\frac{1}{8}}\mathcal{A}^{\frac{3}{2}}\,,
\end{equation}
where $\mathcal{A}$=1.282427129... is Glaisher's constant. % ensuring $\langle\mu\rangle=\bar{s}m^{\frac{1}{8}}$ in the paramagnetic phase. 
Using now that $\mathcal{T}_{1}^{D}(x,0)=\mu(x,0)$, we have 
\begin{equation}
\lim_{\text{QFT}}\langle\mu_{1}^{x}\mu_{j}^{x}\rangle_{{\rm Lat}}=\frac{1}{\bar{s}^{2}J^{\frac{1}{4}}}\langle\mu(0,0)\mu(aj,0)\rangle%\lim_{\text{QFT}}\frac{\bar{s}^{2}J^{\frac{1}{4}}}{\bar{s}^{2}J^{\frac{1}{4}}}\langle\mu_{1}\mu_{L}\rangle=\frac{a^{\frac{1}{4}}}{\bar{s}^{2}2^{\frac{1}{4}}}\langle\mathcal{T}_{1}^{D}(0,0)\tilde{\mathcal{T}}_{1}^{D}(aj,0)\rangle\,,
\label{few}
\end{equation}
The only missing ingredient to find $\zeta_{1}^{D}$ is the relation
between the lattice spacing $a$ and the UV regulator $\varepsilon$
that was established in \cite{ccd-08} and reads 
\begin{equation}
\varepsilon=\chi a,\qquad{\rm with}\qquad\chi=0.0566227\dots.\label{eq:epsilon-a}
\end{equation}
Finally, comparing Eqs. (\ref{LitsOfLimits}) and (\ref{few}), we get 
\begin{equation}
\zeta_{1}^{D}=\frac{1}{\bar{s}^{2}}\left(\frac{2}{\chi}\right)^{\frac{1}{4}}=1.32225\dots\,.\label{eq:Z1D}
\end{equation}
%To have control over $Z_{1}(1)$, the proportionality constant is the only required quantity as two-point function of the Ising QFT
%is known and can be computed in terms of a Painlevé III differential equation \cite{MuVEV}.

An alternative way of calculating $\zeta_{1}^{D}$ consists in taking
the continuum limit of the exact lattice result for the charged moment
$Z_{n}^{(\text{Lat})}(1)$ calculated in Ref. \cite{fg-20} for a
long interval (there it was denoted by $S_{n}^{(-)}$ and was derived
in the XY model, being a generalisation of Ising). In the paramagnetic
phase ($h>1$) in which we are interested, it was
found \cite{fg-20} 
\begin{equation}
\lim_{\ell\to\infty}|Z_{n}^{(\text{Lat})}(1)|=\left[\frac{\left(kk'\right)^{2n}(k'_{n})^{4}}{16^{n-1}k_{n}^{2}}\right]^{\frac{1}{12}},\label{eq:S-nLattice}
\end{equation}
where $k=1/h$, $k'=\sqrt{1-k^{2}}$ and $k_{n}$ and $k'_{n}=\sqrt{1-k_{n}^{2}}$
are the solution of the transcendental equation 
\begin{equation}
\exp\left[-\pi n\frac{I(k')}{I(k)}\right]=\exp\left[-\pi\frac{I(k_{n}')}{I(k_{n})}\right]\,,
\end{equation}
with 
\begin{equation}
I(k)=\int_{0}^{1}\frac{\mathrm{d}x}{\left(1-x^{2}\right)\left(1-k^{2}x^{2}\right)}\,,
\end{equation}
i.e., the complete elliptic integral. Obviously $k_{1}=k$ and $k_{1}'=k'$.
Hence, for $n=1$, Eq. (\ref{eq:S-nLattice}) is just $\lim_{\ell\to\infty}|Z_{1}^{(\text{Lat})}(1)|=\sqrt{k'}$,
that close to the critical point is $\left(2(h-1)\right)^{1/4}=(2ma)^{1/4}$.
On the other hand, directly in the continuum limit we have Eq. \eqref{zn1}, which in the limit of large separation and for $n=1$ is 
\begin{equation}
\lim_{\ell\rightarrow\infty}\zeta_{1}^{D}\varepsilon^{2d_{1}^{D}}\langle\mathcal{T}_{1}^{D}(0,0)\tilde{\mathcal{T}}_{1}^{D}(\ell,0)\rangle=\mathcal{\zeta}_{1}^{D}\varepsilon^{\frac{1}{4}}m^{^{\frac{1}{4}}}\bar{s}^{2},\label{eq:Z1(1)EndResult}
\end{equation}
that provides for $\zeta_{1}^{D}$ exactly the same result as in Eq.
\eqref{eq:Z1D}.

The other amplitude to be calculated is $\frac{\partial\ln\zeta_{n}^{D}}{\partial n}\Big|_{n=1}$
in Eq. \eqref{eq:s(1)}. We can use the last procedure to get this
amplitude using $s^{(\text{Lat})}(1)\equiv-\frac{d}{dn}Z_{n}^{(\text{Lat})}(1)$
derived from Eq. \eqref{eq:S-nLattice} in \cite{fg-20}, obtaining,
for $h>1$, 
\begin{equation}
\lim_{\ell\rightarrow\infty}|s^{(\text{Lat})}(1)|=\frac{\sqrt{k'}}{3}\left[\ln2-\frac{1}{2}\ln\left(kk'\right)-\frac{I(k)I(k')}{\pi}\left(1+k^{2}\right)\right]\,.\label{eq:S-vNLattice}
\end{equation}
Recalling that, by definition, ${\displaystyle \lim_{\text{QFT}}Z_{n}^{(\text{Lat})}(1)=Z_{n}(1)}$,
we have 
\begin{equation}
\lim_{\text{QFT}}Z_{n}^{(\text{Lat})}(1)=\lim_{\ell\rightarrow\infty}\zeta_{n}^{D}\varepsilon^{2d_{n}^{D}}\langle\mathcal{T}_{n}^{D}(0,0)\tilde{\mathcal{T}}_{n}^{D}(\ell,0)\rangle_{n}=\zeta_{n}^{D}\varepsilon^{2d_{n}^{D}}\langle\mathcal{T}_{n}^{D}\rangle^{2}\,.
\end{equation}
Rearranging the previous expression, one can extract $\zeta_{n}^{D}$
and its derivative with respect to $n$ to get 
\begin{equation}
\frac{\text{d}\zeta_{n}^{D}}{\text{d}n}\Big|_{n=1}=\lim_{\text{QFT}}\frac{-s^{(\text{Lat})}(1)}{\varepsilon^{\frac{1}{4}}\langle\mathcal{T}_{1}^{D}\rangle^{2}}-\frac{Z_{1}^{(\text{Lat})}(1)}{\varepsilon^{\frac{1}{2}}\langle\mathcal{T}_{1}^{D}\rangle^{4}}\left(\langle\mathcal{T}_{1}^{D}\rangle^{2}\frac{\text{d}\varepsilon^{2d_{n}^{D}}}{\text{d}n}\Big|_{n=1}+\varepsilon^{\frac{1}{4}}\frac{\text{d}\langle\mathcal{T}_{n}^{D}\rangle^{2}}{\text{d}n}\Big|_{n=1}\right)\,.
\end{equation}
The QFT limit of lattice quantities are simply 
\begin{equation}
\lim_{\text{QFT}}s^{(\text{Lat})}(1)=%\lim_{\deltah\rightarrow0}\left(2\deltah\right)^{\frac{1}{4}}\left(\frac{\ln\deltah}{12}-\frac{\ln2}{4}\right)+o(\deltah^{\frac{1}{4}})=\lim_{a\rightarrow0}
\left(2am\right)^{\frac{1}{4}}\left(\frac{\ln\left(am\right)}{12}-\frac{\ln2}{4}\right)+o(a^{\frac{1}{4}})\,,\label{eq:Leading_s(1)Lattice}
\end{equation}
and 
\begin{equation}
\lim_{\text{QFT}}Z_{1}^{(\text{Lat})}(1)=%\lim_{\deltah\rightarrow0}\left(\deltah\right)^{\frac{1}{4}}+o(\deltah^{\frac{1}{4}})=\lim_{a\rightarrow0}
\left(2am\right)^{\frac{1}{4}}+o(a^{\frac{1}{4}})\,.\label{eq:Leading_Z1(1)Lattice}
\end{equation}
Instead, the VEV $\langle\mathcal{T}_{n}^{D}\rangle^{2}$ and its
derivative are explicitly calculated in appendix \ref{sec:Appendix-A-VEV},
cf. Eqs. \eqref{vev1} and \eqref{vev2}. Putting everything together,
we finally have 
\begin{multline}
\frac{\text{d}\zeta_{n}^{D}}{\text{d}n}\Big|_{n=1}=\lim_{a\to0}\frac{-2^{\frac{1}{4}}\left(\frac{\ln\left(am\right)}{12}-\frac{\ln2}{4}\right)}{\chi^{\frac{1}{4}}\langle m^{-\frac{1}{8}}\mathcal{T}_{1}^{D}\rangle^{2}}-\frac{2^{\frac{1}{4}}}{(ma)^{\frac{1}{4}}\left(\chi^{\frac{1}{4}}\langle m^{-\frac{1}{8}}\mathcal{T}_{1}^{D}\rangle^{2}\right)^{2}}\times\\
\times\left(\langle m^{-\frac{1}{8}}\mathcal{T}_{1}^{D}\rangle^{2}\frac{\text{d}(ma\chi)^{2d_{n}^{D}}}{\text{d}n}\Big|_{n=1}+(ma\chi)^{\frac{1}{4}}\frac{\text{d}\langle m^{-\frac{1}{8}}\mathcal{T}_{n}^{D}\rangle^{2}}{\text{d}n}\Big|_{n=1}\right)\\
=-0.007124\ldots\,.
\end{multline}
Notice that the term in $\ln(am)$ cancels, as it should. We also
used $\varepsilon=a\chi$, cf. Eq \eqref{eq:epsilon-a}.

\subsection{The two-point function of composite twist fields}

Now we change focus and consider the two-point function entering in
Eqs. \eqref{zn1} and \eqref{eq:s(1)}. For $n=1$, the two-point
function of the composite fields in $Z_{1}(1)$ is just
to the two-point function of the disorder operators, which can be
also expressed in terms of a solution of a Painlevé III type differential
equation \cite{MuVEV}. However, for our purposes, the two-particle
approximation of the two-point functions is more useful because it provides 
not only the two-point function at $n=1$, but also its derivative with respect to $n$. In this two-particle approximation,
the correlation function for generic $n$ can be written as (cf. Eq. \eqref{eq:F2TwistFieldSemiLocalZ2}
with \eqref{fminI}) 
\begin{equation}
\begin{split}\langle\mathcal{T}_{n}^{D}(\ell,0)\tilde{\mathcal{T}}_{n}^{D}(0,0)\rangle\approx & \langle\mathcal{T}_{n}^{D}\rangle^{2}+\sum_{j,k=1}^{n}\int_{-\infty}^{\infty}\frac{\mathrm{d}\vartheta_{1}\mathrm{d}\vartheta_{2}}{(2\pi)^{2}2!}|F_{2}^{\mathcal{T}^{D}|j,k}(\vartheta_{12},n)|^{2}e^{-rm\left(\cosh\vartheta_{1}+\cosh\vartheta_{2}\right)}\\
= & \langle\mathcal{T}_{n}^{D}\rangle^{2}\left(1+\frac{n}{4\pi^{2}}\int_{-\infty}^{\infty}\mathrm{d}\vartheta f^{D}(\vartheta,n)K_{0}\left(2m\ell\cosh\left(\vartheta/2\right)\right)\right)\,,
\end{split}
\label{2ptf}
\end{equation}
where $f^{D}(\vartheta,n)$ is implicitly defined as 
\begin{multline}
\langle\mathcal{T}_{n}^{D}\rangle^{2}f^{D}(\vartheta,n)=\sum_{j=1}^{n}|F_{2}^{\mathcal{T}^{D}|1,j}(\vartheta,n)|^{2}=|F_{2}^{\mathcal{T}^{D}|1,j}(\vartheta,n)|^{2}+\sum_{j=1}^{n-1}|F_{2}^{\mathcal{T}^{D}|1,j}(2\pi ij-\vartheta,n)|^{2}\,.\label{eq:f(theta,n)}
\end{multline}
We have already argued that the $k$-particle form factors of the
$\mathbb{Z}_{2}$ twist field vanish for odd $k$ in both the Ising
and ShG models. It has been also shown that the possible presence
of a one-particle FF is irrelevant for the leading behaviour of the
total entropy \cite{cd-09}.
Overall, Eq. \eqref{2ptf} allows us to identify the universal function $H_n(m\ell)$ in Eq. \eqref{zn1} in the two-particle approximation as
\begin{equation}
H^{\rm 2pt}_n(m\ell)=1+\frac{n}{4\pi^{2}}\int_{-\infty}^{\infty}\mathrm{d}\vartheta f^{D}(\vartheta,n)K_{0}(2m\ell\cosh(\vartheta/2))\,,
\label{H2pt}
\end{equation}
an expression that is valid for a generic ${\mathbb Z}_2$ symmetric theory with only the precise form of $f^{D}(\vartheta,n)$ depending on the model. 
Eq. \eqref{H2pt} with \eqref{eq:f(theta,n)} provides an explicit final result for the R\'enyi entropies for any odd integer $n\geq 3$ 
(we recall our FFs are derived for odd $n$). 
The calculation of the von Neumann limit $n\to1$ is more involved because it requires the analytic continuation of Eq. \eqref{eq:f(theta,n)} 
which is not an obvious matter, as we will see soon. However, before embarking in this more difficult calculation, 
let us consider the explicit form of $Z_{1}(1)$. 
In this case, the form factors of the composite twist field become those of the disorder operator, cf. Eq. \eqref{eq:F2DisorderIsing}, 
getting $F_{2}^{\mu}\propto\tanh\vartheta/2$, cf. Eq. \eqref{FmuT}.
Hence we immediately have 
\begin{equation}
H^{\rm 2pt}_1(m\ell)=1+\frac{1}{4\pi^{2}}\int_{-\infty}^{\infty}\mathrm{d}\vartheta\tanh^{2}\Big(\frac{\vartheta}{2}\Big)K_{0}\left(2m\ell\cosh\left(\vartheta/2\right)\right)
= 1+\frac{1}{8\pi}\frac{e^{-2m\ell}}{\left(m\ell\right)^{2}}+\mathcal{O}\Big(\frac{e^{-2m\ell}}{\left(m\ell\right)^{3}}\Big)\,,
\label{eq:T1T1}
\end{equation}
where the leading term in the $m\ell$ expansion is obtained below, but it can also be extracted using
the fact that the integral in \eqref{eq:T1T1} can be rewritten in terms of the Meijer's G-function (although its form is not illuminating and we do not report it here). 

Looking at Eq. \eqref{eq:s(1)} for $s(1)$, we still need the derivative of both the VEV and of the universal function $H^{\rm 2pt}_n(m\ell)$.
The former is rather cumbersome, but does not require any particular care and it is then reported in appendix \ref{sec:Appendix-A-VEV}, see Eq. \eqref{vev2} for the final result.
Conversely, the analytic continuation of $H^{\rm 2pt}_n(m\ell)$ is more thoughtful and we report its details in the following.
In the two-particle approximation, the required derivative reads 
\begin{multline}
\lim_{n\rightarrow1}\frac{\partial}{\partial n}  H^{\rm 2pt}_n(m\ell)=
 \frac{1}{4\pi^{2}}\int_{-\infty}^{\infty}\mathrm{d}\vartheta{\tilde{f}^{D}}(\vartheta,1)K_{0}\left(2\ell m\cosh\left(\vartheta/2\right)\right)\\ 
+  \lim_{n\rightarrow1}\frac{1}{4\pi^{2}}\int_{-\infty}^{\infty}\mathrm{d}\vartheta\left(\frac{\partial}{\partial n}{\tilde{f}^{D}}(\vartheta,n)\right)K_{0}\left(2\ell m\cosh\left(\vartheta/2\right)\right)\,,
\label{eq:-Z1}
\end{multline}
where we introduced $\tilde{f}^{D}(\vartheta,n)$ which is the analytic continuation of $f^{D}(\vartheta,n)$. The evaluation of $\tilde{f}^{D}(\vartheta,1)$
and of its the derivative, nevertheless, involves some subtleties
related to the proper analytic continuation in $n$ of the FFs, which
is non-trivial as carefully discussed in Ref. \cite{ccd-08} for the
conventional twist field. %This procedure is straightforward for the VEV, as the continuation of the formula for the expectation value is very natural; consequently its derivative
%is easy to obtain too (The computation of the VEV and its derivative is indeed presented in appendix \ref{sec:Appendix-A-VEV}). 
%However, the analytic continuation of $f^{D}(\vartheta,n)$ is tricky, similarly to its counterpart for conventional branch-point twist fields \cite{ccd-08}.
For any integer odd $n\geq3$, $\tilde{f}^{D}(\vartheta,n)={f}^{D}(\vartheta,n)$.
This is no longer true for $n=1$: $\tilde{f}^{D}(\vartheta,1)$ is not a continuous function in $\vartheta$, as it equals 
\begin{equation}
\tilde{f}^{D}(\vartheta,1)=\tanh^{2}\frac{\vartheta}{2},
\end{equation}
everywhere except at $\vartheta=0$, where $\tilde{f}^{D}(0,1)=-\frac{1}{2}$.
In other words, $\tilde{f}^{D}(\vartheta,1)$ equals $f^{D}(\vartheta,1)$
everywhere, except at $\vartheta=0$. Consequently, its derivative contains a $\delta$-function. 
The calculation is detailed in appendix \ref{sec:Appendix-C-AnalContForfD}, where one finally arrives to Eq. \eqref{Ancon}, i.e., 
\begin{equation}
\lim_{n\to1}\frac{\partial}{\partial n}\tilde{f}^{D}(\vartheta,n)=\frac{1}{2}\frac{1-\cosh\vartheta+\frac{2\vartheta}{\sinh\vartheta}}{\cosh^{2}\frac{\vartheta}{2}}-\pi^{2}\frac{1}{2}\delta(\vartheta) = 4 \vartheta \frac{\sinh^2(\vartheta/2)}{\sinh^3\vartheta}-\tanh^2(\vartheta/2)-\pi^{2}\frac{1}{2}\delta(\vartheta)  \,,
\end{equation}
It follows that the final result for Eq. \eqref{eq:-Z1} is
\begin{equation}
\lim_{n\rightarrow1}\frac{\partial}{\partial n}  H^{\rm 2pt}_n(m\ell)=
\frac1{\pi^2} \int_{-\infty}^\infty  \mathrm{d}\vartheta  \frac{\vartheta\sinh^2(\vartheta/2)}{\sinh^3\vartheta}K_{0}\left(2\ell m\cosh\left(\vartheta/2\right)\right)-\frac18 K_0(2 m\ell)
\,,
\label{eq:-Z11}
\end{equation}
This term, together with \eqref{eq:T1T1} includes the entire $\ell$ dependence of the symmetry resolved von Neumann entropies and 
it represents our final full result. 

However, putting the various pieces together is not illuminating without expanding for large $m\ell$ as we are going to do now. 
The leading term in \eqref{eq:-Z11} clearly comes from the $K_0(m\ell)$ factor, but it is worth discussing a simple method to obtain a
systematic large $\ell$ expansion. 
%As a consequence, the first integral in Eq. (\ref{eq:-Z1}), the one involving $\tilde{f}^{D}(\vartheta,1)$, is of order $\mathcal{O}(\frac{e^{-2m\ell}}{\left(m\ell\right)^{2}})$
%while the second one, containing $\frac{\partial}{\partial n}\tilde{f}^{D}(\vartheta,n)$, has both $\mathcal{O}(\frac{e^{-2m\ell}}{m\ell})$ and $\mathcal{O}(e^{-2m\ell})$
%parts, with the latter originating from the $\delta$-function. 
To obtain the subleading terms by evaluating the integrals in Eqs. \eqref{eq:-Z11} and \eqref{eq:T1T1}, one first recognises that  for large $\ell$, the integral is
dominated by the contribution of the region close to $\vartheta=0$. 
One can then expand as a function of $\vartheta=0$ the function which multiply $K_0(m\ell)$ in the integrand, and evaluate the asymptotic behaviour of  
\begin{equation}
\frac{1}{4\pi^{2}}\int_{-\infty}^{\infty}\text{d}\vartheta K_{0}(2m\ell\cosh\frac{\vartheta}{2})\left(\frac{\vartheta}{2}\right)^{2n}
=
\frac{1}{\pi^{2}}\int_{1}^{\infty}\text{d} x \frac{{\rm arccosh}^{2n} x}{\sqrt{x^2-1}}K_{0}(2m\ell x)
\,.\label{eq:BesselTheta2n}
\end{equation}
Expanding $\text{arcosh}(x)$ around $x=1$, exploiting the asymptotic behaviour  of the Bessel function
$K_{0}(z)\approx e^{-z}\sqrt{\frac{\pi}{2z}}$, and keeping the leading $x-1$ type terms, we and up with  
\begin{equation}
\frac{1}{\pi^{2}}\int_{1}^{\infty}\text{d}xe^{-2m\ell x}\sqrt{\frac{\pi}{4m\ell}}\frac{2^{n}\sqrt{x-1}^{2n-1}}{\sqrt{2}}=
\frac{\Gamma\left(n+\frac{1}{2}\right)}{4\pi^{3/2}} \frac{e^{-2m\ell}}{(m\ell)^{n+1}} (1+ O((m\ell)^{-1})\,,
\end{equation}
which gives the leading $\ell$-dependent term for \eqref{eq:BesselTheta2n}.
In this way, one readily derive the expansion in the rhs of Eq. \eqref{eq:T1T1} and  
\begin{equation}
\lim_{n\rightarrow1}\frac{\partial}{\partial n}  H^{\rm 2pt}_n(m\ell)=
-\frac18 K_0(2m\ell)+\frac1{4\pi} \frac{e^{-2m\ell}}{m\ell} %-\frac5{96 \pi}  \frac{e^{-2m\ell}}{(m\ell)^2} 
+O\Big( \frac{e^{-2m\ell}}{(m\ell)^2}\Big).
\label{dH1exp}
\end{equation}

%The analytic part of $\lim_{n\rightarrow1}\frac{\partial}{\partial n}\tilde{f}^{D}(\vartheta,n)$
%behaves as $1-\mathcal{O}(\vartheta^{2})$ and neglecting the discontinuity
%of $\tilde{f}^{D}(\vartheta,n)$, we have $\left(\frac{\vartheta}{2}\right)^{2}$
%for small $\vartheta$. This way, we can write 
%\begin{multline}
%\lim_{n\rightarrow1}\frac{\partial}{\partial n}m^{-2d_{n}^{D}}\langle\mathcal{T}_{n}^{D}(0,0)\tilde{\mathcal{T}}_{n}^{D}(\ell,0)\rangle\approx\lim_{n\rightarrow1}\frac{\mathrm{d}}{\mathrm{d}n}\left(m^{-2d_{n}^{D}}\langle\mathcal{T}_{n}^{D}\rangle^{2}\right){\color{red}\left(1+\frac{1}{8\pi}\frac{e^{-2m\ell}}{\left(m\ell\right)^{2}}+\mathcal{O}(\frac{e^{-2m\ell}}{\left(m\ell\right)^{3}})\right)}\\
%{\color{black}{\color{red}-}{\color{black}{\color{red}m^{-2d_{n}^{D}}\langle\mathcal{T}_{n}^{D}\rangle^{2}\frac{1}{8}K_{0}\left(2\ell m\right)+m^{-2d_{n}^{D}}\langle\mathcal{T}_{n}^{D}\rangle^{2}\frac{1}{4\pi}\frac{e^{-2m\ell}}{m\ell}+\mathcal{O}(\frac{e^{-2m\ell}}{\left(m\ell\right)^{2}})\,.}}}
%\end{multline}

\subsection{Putting the pieces together}

In this subsection we put together the different pieces of the symmetry resolved entropies. We first of all write down the
expressions for $Z_{1}(1)$ and $s(1)$ including the leading corrections and then comment on the symmetry resolved entropy.  
$Z_{1}(1)$ is obtained by plugging Eqs. \eqref{eq:T1T1} and \eqref{eq:Z1(1)EndResult} into Eq. \eqref{zn1}, getting  
\begin{equation}
Z_{1}(1)=\zeta_{1}^{D}(m\varepsilon)^{\frac{1}{4}}\bar{s}^{2}\left(1+\frac{1}{8\pi}\frac{e^{-2m\ell}}{\left(m\ell\right)^{2}}+\mathcal{O}(\frac{e^{-2m\ell}}{\left(m\ell\right)^{3}})\right)\,,\label{z11Final}
\end{equation}
 $\bar{s}=2^{\frac{1}{12}}e^{-\frac{1}{8}}\mathcal{A}^{\frac{3}{2}}$ and $\zeta_{1}^{D}=1.32225\dots\,$, as obtained in Sec. \ref{sec:ampl}.  
In a similar fashion, $s(1)$ is obtained by plugging Eqs. \eqref{dH1exp}, \eqref{eq:T1T1} into \eqref{eq:s(1)}, getting
\begin{multline}
s(1)=-\zeta_{1}^{D}(m\varepsilon)^{\frac{1}{4}}\bar{s}^{2}\left(1+\frac{1}{8\pi}\frac{e^{-2m\ell}}{\left(m\ell\right)^{2}}+\mathcal{O}(\frac{e^{-2m\ell}}{\left(m\ell\right)^{3}})\right)
\\ \times
\left[-\frac{\ln m\varepsilon}{12}+{\cal C}
-\frac{1}{8}K_{0}\left(2\ell m\right)+\frac{1}{4\pi}\frac{e^{-2m\ell}}{m\ell}+\mathcal{O}\Big(\frac{e^{-2m\ell}}{\left(m\ell\right)^{2}}\Big)\right]\,,
\label{eq:S-IsingLeadingQFTA}
\end{multline}
where we introduced the combination of amplitudes
\begin{equation}
\mathcal{C}=\lim_{n\rightarrow1}\left(\frac{\text{d}\ln\zeta_{n}^{D}}{\text{d}n}+\frac{\text{d}}{\text{d}n}\ln\left(m^{-2d_{n}^{D}}\langle\mathcal{T}_{n}^{D}\rangle^{2}\right)\right)=-0.065992\,,
\end{equation}
with the numerical value coming from  $\displaystyle\lim_{n\rightarrow1}\frac{\text{d}\ln\zeta_{n}^{D}}{\text{d}n}=-0.00538786$
and $\displaystyle\lim_{n\rightarrow1}\frac{\text{d}}{\text{d}n}\ln\left(m^{-2d_{n}^{D}}\langle\mathcal{T}_{n}^{D}\rangle^{2}\right)=-0.0606041$,
as calculated in Sec. \ref{sec:ampl}.  
Slightly rephrasing the formula using $\varepsilon=\chi a$, we have
\begin{multline}
s(1)=  \left(2am\right)^{\frac{1}{4}}\left(1+\frac{1}{8\pi}\frac{e^{-2m\ell}}{\left(m\ell\right)^{2}}+\mathcal{O}(\frac{e^{-2m\ell}}{\left(m\ell\right)^{3}})\right)
\\ \times
\left[\left(\frac{\ln\left(am\right)}{12}+\frac{\ln\chi}{12}-\mathcal{C}\right)+\frac{1}{8}K_{0}\left(2\ell m\right)+-\frac{1}{4\pi}\frac{e^{-2m\ell}}{m\ell}+\mathcal{O}(\frac{e^{-2m\ell}}{\left(m\ell\right)^{2}})\right]\,,
\label{eq:S-IsingLeadingQFT}
\end{multline}
which can be cross-checked against the lattice result \eqref{eq:Leading_s(1)Lattice}.
The equality of $-\frac{\ln2}{4}$ in \eqref{eq:Leading_s(1)Lattice}
and $\frac{\ln\chi}{12}-\mathcal{C}$ can be regarded as a consistency check of the calculations. 
In our results for $s(1)$ i.e., in Eqs. \eqref{eq:S-IsingLeadingQFTA} and \eqref{eq:S-IsingLeadingQFT} we also kept the leading and subleading terms accounting for the $\ell$-dependence. 
The analogous term incorporating $\ell$-dependence has not been derived for the lattice model and 
represent one of our main achievements.

With \eqref{z11Final} for $Z_{1}(1)$ and \eqref{eq:S-IsingLeadingQFTA} for $s(1)$, we can finally use \eqref{eq:S0S1exp} to
write down the symmetry resolved entropies including corrections too.
Keeping the $\varepsilon^{1/4}\ln\varepsilon$ and $\varepsilon^{1/4}$ terms, we end up with
\begin{equation}
\begin{split}S(\pm)= & -\frac{1}{6}\ln m\varepsilon+U_{Ising}-\frac{1}{8}K_{0}(2m\ell)-\ln2\pm\left(\frac{2}{\chi}\right)^{\frac{1}{4}}\left(\varepsilon m\right)^{\frac{1}{4}}\left(1+\frac{1}{8\pi}\frac{e^{-2m\ell}}{\left(m\ell\right)^{2}}\right)\left[\frac{\ln\left(\varepsilon m\right)}{4}+\right.\\
 & \left.\qquad\qquad-U_{Ising}-\mathcal{C}+\frac{1}{4}K_{0}\left(2m\ell\right)-\frac{1}{4\pi}\frac{e^{-2m\ell}}{m\ell}\right]
 +\mathcal{O}\Big(e^{-3m\ell},\varepsilon^{\frac{1}{4}}\ln\varepsilon\frac{e^{-2m\ell}}{\left(m\ell\right)^{3}},\varepsilon^{\frac{1}{4}}\frac{e^{-2m\ell}}{\left(m\ell\right)^{2}}\Big)\\
= & -\frac{1}{6}\ln m\varepsilon-0.131984-\frac{1}{8}K_{0}(2m\ell)-\ln2\pm2.437866\left(\varepsilon m\right)^{\frac{1}{4}}\left[\frac{\ln\left(\varepsilon m\right)}{4}\left(1+\frac{1}{8\pi}\frac{e^{-2m\ell}}{\left(m\ell\right)^{2}}\right)+\right.\\
 & \left.\qquad\qquad+0.197976+\frac{1}{4}K_{0}\left(2m\ell\right)-\frac{1}{4\pi}\frac{e^{-2m\ell}}{m\ell}\right]
 +\mathcal{O}\Big(e^{-3m\ell},\varepsilon^{\frac{1}{4}}\ln\varepsilon\frac{e^{-2m\ell}}{\left(m\ell\right)^{3}},\varepsilon^{\frac{1}{4}}\frac{e^{-2m\ell}}{\left(m\ell\right)^{2}}\Big)\,.
\end{split}
\label{eq:S0S1RDepPlusCorr}
\end{equation}
As already anticipated on a general ground in Sec. \ref{sec:gen} Eq. \eqref{finalEQ}, we find at leading order equipartition of entanglement, 
i.e. $S(+)=S(-)+\dots$. On top of this, the above expression can be used to find the first term breaking equipartition which can be easily extracted by 
taking the difference
\begin{multline}
\frac{S(+)-S(-)}2= 2.437866\left(\varepsilon m\right)^{\frac{1}{4}}\left[\frac{\ln\left(\varepsilon m\right)}{4}\left(1+\frac{1}{8\pi}\frac{e^{-2m\ell}}{\left(m\ell\right)^{2}}\right)+
0.197976+\frac{1}{4}K_{0}\left(2m\ell\right)-\frac{1}{4\pi}\frac{e^{-2m\ell}}{m\ell}\right]\\
 +\mathcal{O}\Big(e^{-3m\ell},\varepsilon^{\frac{1}{4}}\ln\varepsilon\frac{e^{-2m\ell}}{\left(m\ell\right)^{3}},\varepsilon^{\frac{1}{4}}\frac{e^{-2m\ell}}{\left(m\ell\right)^{2}}\Big)
 \label{diffS}
\end{multline}
 It should be possible to test this prediction by exact numerical lattice computation. 
 Work in this direction is in progress.

\section{Conclusions\label{sec:Conclusions}}

In this paper, we introduced an approach suited to the computation of symmetry resolved entropies in generic massive (free and interacting)
integrable quantum field theories. 
The essence of the approach is the existence of appropriate modified or composite branch-point twist
fields whose two-point function gives the corresponding charged entropies for a single interval. 
Then the form factor bootstrap program provides the matrix elements of such fields.
In particular, here we discussed the $\mathbb{Z}_{2}$ symmetry resolution for Ising model in the paramagnetic phase and for the sinh-Gordon
quantum field theory.

%For a discrete symmetry such as the $\mathbb{Z}_{2}$ case, the validity of the symmetry resolution of entropies is not obvious for any subsystem,
%even if the total system is an eigenstate of the symmetry. We therefore explicitly showed that a $\mathbb{Z}_{2}$ symmetry resolutions is
%sensible in the Ising model if the subsystem is an interval and provided arguments for the validity of the resolution in the sinh-Gordon model
%as long as its ground state is considered and the subsystem is an interval. 
We wrote down the bootstrap equations for the composite twist fields and provided an intuitive picture behind the choice
of the locality factors entering these equations. 
The two-particle form factors for $\mathbb{Z}_{2}$ branch-point twist fields were calculated for the Ising both models considered here. 
For the Ising model, we were also able to compute the vacuum expectation value, alias the zero particle form factor, 
we argued that form factors with odd particle number vanish, and finally showed that  the form factors for any even particle numbers can
are Pfaffian of the two-particle form factors. 
The obtained form factor solution was cross-checked verifying that for $n\to1$ the form factors of the disorder operator are recovered and applying 
the $\Delta$-theorem \cite{delta_theorem} to reproduce exactly the critical dimensions of the composite fields. 

Also the sinh-Gordon form factors have been tested in several ways.
First, we considered the limit for the interaction parameter $B$ as $B=1+i\frac{2}{\pi}\Theta_{0}$
with $\Theta_{0}\rightarrow\infty$, in which the $\mathbb{Z}_{2}$ branch-point twist fields for the Ising model are recovered.
Then for $n\rightarrow1$, we reproduced the disorder operator of the sinh-Gordon model. 
Applying the $\Delta$-theorem for the form factors, we recovered the expected UV dimensions with satisfactory precision. 
The error is ascribed to the fact that, unlike for the Ising model, the $\Delta$-theorem sum
rule requires an infinite summation and hence the knowledge of all form factors for the $\mathbb{Z}_{2}$ branch-point twist field.

The general approach to extract the ground-state symmetry resolved entropies for an interval of length $\ell$ from the two-point function
of composite twist fields is discussed in Sec. \ref{sec:gen}. 
In particular, we showed that entanglement equipartition follows generically from the property that the UV dimension of the composite twist field
is larger than the one for the conventional twist field. 
The subleading term breaking such equipartition is model dependent. 
The obtained form factors allow for the complete calculation of the charged and symmetry resolved entropies in the paramagnetic phase of the Ising model which 
is presented in great detail, with emphasis on the physically relevant von Neumann limit $n\to1$ (that requires a non-trivial analytic continuation).
The final result for the charged partition sum and entropy are reported in Eqs. \eqref{zn1} and \eqref{eq:s(1)} with the various amplitudes computed 
in Sec. \ref{sec:ampl} and the universal functions of $m\ell$ given in Eqs. \eqref{eq:T1T1} and \eqref{eq:-Z11}. 
We stress that these universal functions are the main new physical results of this paper since all other terms could be equivalently calculated 
by taking the continuum limit of the known results for the Ising chain in Ref. \cite{fg-20}. 
From Eq. \eqref{diffS} we can see that the leading term breaking equipartition scales like $\varepsilon^{\frac{1}{4}}\ln\varepsilon$, as expected. 
However, Eq. \eqref{diffS} also provides the $m\ell$ dependence of this equipartition breaking term. 
It would be highly desirable to test all these predictions with exact numerical calculations based on the continuum limit of the spin chain.

There are various possible ways this work can be extended. The most natural one is the treatment of models with non-diagonal scattering
and continuous symmetries, to which the authors plan to devote another communication. 
The obtained form factors also allow for the calculation of entropies in excited states, as long as reduced density matrix commutes with the symmetry operator. 
Finally, the crossover from critical to massive regime at fixed $\ell$ is a very interesting yet challenging problem,
which may require an infinite summation higher particle form factors or the development of alternative techniques.

\subsection*{Acknowledgments}
DXH is grateful to Sara Murciano for many useful discussions.
The authors are also grateful to Olalla Castro-Alavaredo for fruitful feedbacks on a first version of the manuscript. 
PC and DXH acknowledge support from ERC under Consolidator grant number 771536 (NEMO).

\appendix

\section{Vacuum expectation value of $\mathcal{T}_{\text{Ising}}^{D}$\label{sec:Appendix-A-VEV}}

Finding the solutions to the FF bootstrap equations is relatively
easy. Often it is also not difficult to identify these solutions with
the corresponding physical fields. Conversely, the determination of
the vacuum expectation value (VEV), i.e., the zero particle FF and
the one-particle FF (if non-vanishing) is generally a difficult task.
So far, exact expressions are known for all fields in the Ising model
and for some in ShG, sine-Gordon, Bullogh-Dodd models, as well as
for some of their restrictions, see e.g. \cite{m-book,VEV1,VEV2,VEV3}.
For the conventional branch-point twist fields, an exact expression
for the VEV has been provided only for the Ising model in \cite{ccd-08}.
In this appendix, we show that for the same model the VEV for $\mathcal{T}_{n}^{D}$
can also be exactly determined, under some plausible assumptions.
We use and modify ideas borrowed from Refs. \cite{ccd-08,cs-17,cfh-07}.
In this appendix, we work in the fermionic basis and denote the $j$-th
copy of the Majorana fermion as $\psi_{j}$. We explicitly exploit the property that 
fermionic and spin entanglement are the same for one interval.  

As a first step we search for a matrix $\tau$ whose action in the
space replica space (i.e. on the vector $\left(\psi_{1}^{\text{}},...,\psi_{n}^{\text{}}\right)^{T}$)
corresponds to the the composite twist field. Given that the total
phase accumulated by the field in turning around the entire Riemann
surface is $-1$, the main requirement is $\tau^{n}\psi_{j}=-\psi_{j}$,
i.e., $\tau^{n}=-\mathcal{I}$, where $\mathcal{I}$ is the $n\times n$
identity matrix. An easy way to proceed is to modify the transformation
matrix for the conventional twist-fields \cite{cfh-07}, as done in
Ref. \cite{mdc-20b} for the resolution of the $U(1)$ symmetry (both
papers consider Dirac fermions, but there is no difference for Majorana
except that the phase is fixed). Hence, a first representation of
the matrix $\tau$ is 
\begin{equation}
\tau_{1}=\left(\begin{array}{ccccccc}
0 & 0 & 0 & 0 & \cdots & 0 & (-1)^{n}\\
-1 & 0 & 0 & 0 & \cdots & 0 & 0\\
0 & -1 & 0 & 0 & \cdots & 0 & 0\\
0 & 0 & -1 & 0 &  & 0 & 0\\
\vdots & \vdots &  & \ddots & \ddots &  & \vdots\\
0 & 0 & 0 & 0 & \ddots & 0 & 0\\
0 & 0 & 0 & 0 & \cdots & -1 & 0
\end{array}\right)\label{eq:sigmaMatrixCassiniHuerta(-1)}
\end{equation}
where it is clear that $\tau_{1}^{n}=-\mathcal{I}$ for odd $n$.
However, it was pointed out in \cite{ccd-08} that one has to be careful
in the FF approach because fermions of the same copy anticommute,
as conventional fermions do, but the fermions of different copies
commute ($S_{ij}=1$). Conversely, in Refs. \cite{mdc-20b,cfh-07}
fermions of different copies anticommute. The anticommutation of fermions
on different copies can be achieved in the FF approach by a change
of basis as \cite{ccd-08} 
\begin{equation}
|\vartheta_{1},\vartheta_{2}\rangle_{j_{1},j_{2}}^{\text{ac}}=\begin{cases}
|\vartheta_{1},\vartheta_{2}\rangle_{j_{1},j_{2}} & j_{1}\leq j_{2},\\
-|\vartheta_{1},\vartheta_{2}\rangle_{j_{1},j_{2}} & j_{1}>j_{2}\,.
\end{cases}\label{eq:AC-Basis}
\end{equation}
As argued in \cite{ccd-08}, the action of a permutation on the fields
$\psi_{j}^{\text{ac}}$ in the new basis is no longer $\sigma\psi_{j}^{\text{ac}}=\psi_{j+1\text{ mod }n}^{\text{ac}}$,
but instead 
\begin{equation}
\sigma\psi_{j}^{\text{ac}}=\begin{cases}
\psi_{j+1}^{\text{ac}} & j=1,...,n-1,\\
-\psi_{1}^{\text{ac}} & j=n\,.
\end{cases}\label{sigmaPsiAC}
\end{equation}
When this permutation is applied $n$ times we have $\sigma^{n}\psi_{j}^{\text{ac}}=-\psi_{j}^{\text{ac}}$.
Moreover, the eigenvalues of the corresponding matrix 
\begin{equation}
\tau_{2}=\left(\begin{array}{ccccccc}
0 & 0 & 0 & 0 & \cdots & 0 & -1\\
1 & 0 & 0 & 0 & \cdots & 0 & 0\\
0 & 1 & 0 & 0 & \cdots & 0 & 0\\
0 & 0 & 1 & 0 &  & 0 & 0\\
\vdots & \vdots &  & \ddots & \ddots &  & \vdots\\
0 & 0 & 0 & 0 & \ddots & 0 & 0\\
0 & 0 & 0 & 0 & \cdots & 1 & 0
\end{array}\right)\label{eq:sigmaDoyon}
\end{equation}
equal those of (\ref{eq:sigmaMatrixCassiniHuerta(-1)}) for odd $n$, which the case we are interested in. 
We can then identify both $\tau_{2}$ and $\tau_{1}$ with the transformation matrix that has to be diagonalised
for the determination of the VEV \cite{ccd-08}.

The eigenvalues of $\tau_{1,2}$ can be written as $e^{i2\pi k/n}$
with $k$ 
\begin{equation}
k=-(n-2)/2,-(n-4)/2\ldots,-1/2,1/2,\ldots,(n-4)/2,(n-2)/2,n/2\,.
\end{equation}
The eigenvectors of $\tau_{2}$ are 
\begin{equation}
\psi_{k}=\frac{1}{\sqrt{n}}\sum_{j=1}^{n}e^{-2\pi ik(j-1)/n}\psi_{j}^{ac}\,,
\end{equation}
and the inverse transformation is 
\begin{equation}
\psi_{j}^{ac}=\frac{1}{\sqrt{n}}\sum_{k=-\frac{n-2}{2}}^{\frac{n}{2}}e^{2\pi ik(j-1)/n}\psi_{k}\,.
\end{equation}
The eigenvectors corresponding to the eigenvalues $e^{i2\pi k/n}$
are complex conjugate pairs for $\pm k$, except $k=n/2$ with eigenvalue
$(-1)$ and real eigenvector equal to $\frac{1}{\sqrt{n}}(1,-1,1,...,1)$.
Hence, we can build $\frac{n-1}{2}$ complex fermions by $\psi_{k}$
and $\psi_{-k}$ as $\psi_{k}^{\dagger}=\psi_{-k}$ for $k=1,\ldots,(n-2)$
and we are left with one Majorana fermion for $k=n/2$, which is still
a Majorana fermion as $\psi_{n/2}^{\dagger}=\psi_{n/2}$. The anticommutation
relations $\{\psi_{k},\psi_{k'}\}=\delta_{k,-k'}$, $\{\psi_{k},\psi_{n/2}\}=0$
for $k\neq n/2$, and $\{\psi_{n/2},\psi_{n/2}\}=1$ are ensured by
our choice for the basis (\ref{eq:AC-Basis}).

The structure of the eigenvalues of the transformation $\tau$ is
compatible with the four-point function of the $\mathbb{Z}_{2}$ twist
field 
\begin{equation}
\frac{\langle\psi_{-k}(z)\psi_{k}(z')\mathcal{T}_{n}^{D}(w)\tilde{\mathcal{T}}_{n}^{D}(w')\rangle}{\langle\mathcal{T}_{n}^{D}(w)\tilde{\mathcal{T}}_{n}^{D}(w')\rangle}=\frac{1}{z-z'}\left(\frac{\left(z-w\right)\left(z'-w'\right)}{\left(z-w'\right)\left(z'-w\right)}\right)^{\frac{k}{n}}\,,\label{eq:ComplexFermionDublet4ptFunctionZ2}
\end{equation}
at the UV critical point: turning clock-wise $\psi_{k}(z')$ around the twist field $\mathcal{T}^{D}$ at $w$, the correct factor of
$e^{i2\pi k/n}$ is recovered. Eq. (\ref{eq:ComplexFermionDublet4ptFunctionZ2})
is an important formula, which is also proved in Appendix \ref{sec:Appendix-B-CFTDimensions}.
It leads to the factorisation of the $\mathbb{Z}_{2}$ branch-point
twist field, it allows for the computation of the UV dimensions of
the factorised components, and eventually it leads to the determination
of the VEV in the massive theory. The factorisation of the $\mathbb{Z}_{2}$
twist field can also be inferred from the results of \cite{cs-17}, which in our case become 
\begin{equation}
\mathcal{T}_{n}^{D}(w)=\mathcal{T}_{\frac{n}{2},n}^{D}(w)\prod_{k\geq\frac{1}{2}}^{\frac{n-2}{2}}\mathcal{T}_{k,n}^{D}(w)\,,
\end{equation}
where action of $\mathcal{T}_{k,n}^{D}(w)$ is non trivial only on
the $\psi_{-k}$ and $\psi_{k}$ fields. The scaling dimension of
$\mathcal{T}_{k,n}^{D}$ can be can be obtained from the relation \cite{dixon,k-87,cc-04}
\begin{equation}
\frac{\langle T_{k}(z)\mathcal{T}_{k,n}^{D}(w)\tilde{\mathcal{T}}_{k,n}^{D}(w')\rangle}{\langle\mathcal{T}_{k,n}^{D}(w)\tilde{\mathcal{T}}_{k,n}^{D}(w')\rangle}=h_{k}\frac{\left(w-w'\right)^{2}}{\left(z-w\right)^{2}\left(z-w'\right)^{2}}\,,\label{eq:TTDTDTilde}
\end{equation}
where $T_{k}$ is the stress-energy tensor of the $\pm k$ components.
In fact, using the Ward identity \cite{WardIdentity} 
\begin{equation}
\langle T_{k}(z)\mathcal{T}_{k,n}^{D}(w)\tilde{\mathcal{T}}_{k,n}^{D}(w')\rangle=\left(\frac{\partial_{w}}{z-w}+\frac{h_{\mathcal{T}_{k}}}{\left(z-w\right)^{2}}+\frac{\partial_{w'}}{z-w'}+\frac{h_{\mathcal{\tilde{T}}_{k}}}{\left(z-w'\right)^{2}}\right)\langle\mathcal{T}_{k,n}^{D}(w)\tilde{\mathcal{T}}_{k,n}^{D}(w')\rangle\,,\label{WI}
\end{equation}
one can deduce that the coefficient $h_{k}$ in (\ref{eq:TTDTDTilde})
equals the conformal dimension of the chiral component of both $\mathcal{T}_{n}^{D}$
and $\tilde{\mathcal{T}}_{n}^{D}$.

To calculate (\ref{eq:TTDTDTilde}), we first show, that the stress-energy
tensor can also be factorised into different $k$-components. We recall
that the 2D free massless Dirac theory can be written in terms of
the two component Dirac spinor $\Psi(z,\bar{z})=\binom{\chi(z)}{\bar{\chi}(\bar{z})}$,
where $\chi$ and $\bar{\chi}$ are complex fermion fields. The analytic
part of the stress energy tensor is 
\begin{equation}
T_{\text{Dirac}}(z)=\frac{1}{2}\left(\partial_{z}\Psi^{\dagger}\Psi-\Psi^{\dagger}\partial_{z}\Psi\right)=\frac{1}{2}\left(\partial_{z}\left(\chi{}^{\dagger}(z)\chi(z)\right)-\chi^{\dagger}(z),\partial_{z}\chi(z)\right),\label{eq:DiracStressETensor}
\end{equation}
whereas for the neutral Majorana field it reads 
\begin{equation}
T_{\text{Majorana}}(z)=-\frac{1}{2}\psi(z)\partial_{z}\psi(z)\,.
\end{equation}
One Dirac field can be constructed from two Majorana fields as 
\begin{equation}
\Psi(z,\bar{z})=\binom{\chi(z)}{\bar{\chi}(\bar{z})}=\frac{1}{\sqrt{2}}\binom{\psi_{1}(z)+i\psi_{2}(z)}{\bar{\psi}_{1}(\bar{z})+i\bar{\psi}_{2}(\bar{z})}\,,
\end{equation}
but in our case, as argued before, it is more convenient to use 
\begin{equation}
\Psi_{k}(z,\bar{z})=\binom{\chi_{k}(z)}{\bar{\chi}_{k}(\bar{z})}=\frac{1}{\sqrt{2}}\binom{\psi_{k}(z)}{\bar{\psi}_{k}(\bar{z})},
\end{equation}
with our Fourier transformed fields $\psi_{k}$. % Indeed, the known anti-commutation relations for Dirac fields are recovered. 
In this way, the stress-energy tensor of the original $n$-copy model
is decomposed into $k$ sectors each involving complex fermion fields.
Using Eq. (\ref{eq:DiracStressETensor}), the stress-energy tensor
of the $\pm k$ components is %on the fact that $\psi_{k}^{\dagger}=\psi_{k}^{*}$, 
\begin{equation}
T_{k}=\frac{1}{2}\left(\partial_{z}\psi_{k}^{\dag}\psi_{k}-\psi_{k}^{\dag}\partial_{z}\psi_{k}\right)\,,
\end{equation}
for $k=\frac{1}{2},\ldots,\frac{n-2}{2}$ and, similarly for $k=\frac{n}{2}$
\begin{equation}
T_{\frac{n}{2}}=-\frac{1}{2}\left(\psi_{\frac{n}{2}}\partial_{z}\psi_{\frac{n}{2}}\right)\,.
\end{equation}
The total stress-energy tensor is then 
\begin{equation}
\sum_{k=\frac{1}{2}}^{\frac{n}{2}}T_{k}=\sum_{j=1}^{n}-\frac{1}{2}\left(\psi_{j}\partial_{z}\psi_{j}\right).
\end{equation}

Now we explicitly compute the lhs. of Eq. (\ref{eq:TTDTDTilde}) to
determine $h_{k}$. We first notice that the action of 
\begin{equation}
\frac{1}{2\pi i}\oint\frac{\text{d}z'}{z'-z}\left(-\frac{1}{2}\left[\partial_{z'}-\partial_{z}\right]\right)\,,\label{ResidueApplyDirac}
\end{equation}
to the lhs of Eq. (\ref{eq:ComplexFermionDublet4ptFunctionZ2}) replaces
$\psi_{-k}(z)\psi_{k}(z')$ with $T_{k}(z)$. The operator (\ref{ResidueApplyDirac})
is straightforwardly applied to the rhs of Eq. (\ref{eq:ComplexFermionDublet4ptFunctionZ2})
and so the scaling dimension $h_{k}$ is 
\begin{equation}
h_{k}=\frac{k^{2}}{2n^{2}},
\end{equation}
for $k=\frac{1}{2},\ldots,\frac{n-2}{2}$. Finally $\mathcal{T}_{\frac{n}{2},n}(w,\bar{w})$
acts like the conventional disorder operator and so 
\begin{equation}
h_{\frac{n}{2}}=\frac{1}{16}\,.
\end{equation}
This dimension can be also rigorously obtained by applying 
\begin{equation}
\frac{1}{2\pi i}\oint\frac{\text{d}z'}{z'-z}\left(-\frac{1}{4}\left[\partial_{z'}-\partial_{z}\right]\right)\,,\label{ResidueApplyMajorana}
\end{equation}
to 
\begin{equation}
\frac{\langle\psi_{\frac{n}{2}}(z)\psi_{\frac{n}{2}}(z')\mathcal{T}_{n}^{D}(w)\tilde{\mathcal{T}}_{n}^{D}(w')\rangle}{\langle\mathcal{T}_{n}^{D}(w)\tilde{\mathcal{T}}_{n}^{D}(w')\rangle}=\frac{1}{z-z'}\left(\frac{\left(z-w\right)\left(z'-w'\right)}{\left(z-w'\right)\left(z'-w\right)}\right)^{\frac{1}{2}}\,.\label{eq:MajoranaFermionDoublet}
\end{equation}
The factor $\frac{1}{4}$ in (\ref{ResidueApplyMajorana}) compared
to $\frac{1}{2}$ in (\ref{ResidueApplyDirac}) is important to obtain
the desired $-\frac{1}{2}\psi_{\frac{n}{2}}(z)\partial_{z}\psi_{\frac{n}{2}}(z)$
with the correct normalisation. The application of (\ref{ResidueApplyMajorana})
to (\ref{eq:ComplexFermionDublet4ptFunctionZ2}) results in 
\begin{equation}
\frac{\langle T_{\frac{n}{2}}(z)\mathcal{T}_{\frac{n}{2},n}^{D}(w)\tilde{\mathcal{T}}_{\frac{n}{2},n}^{D}(w')\rangle}{\langle\mathcal{T}_{\frac{n}{2},n}^{D}(w)\tilde{\mathcal{T}}_{\frac{n}{2},n}^{D}(w')\rangle}=\frac{1}{16}\frac{\left(w-w'\right)^{2}}{\left(z-w\right)^{2}\left(z-w'\right)^{2}}
\end{equation}
confirming $h_{\frac{n}{2}}=\frac{1}{16}$.

Finally, the total dimension of the composite twist field is 
\begin{equation}
\frac{1}{2}\sum_{k=\frac{1}{2}}^{\frac{n-2}{2}}\frac{k^{2}}{2n}+\frac{1}{16}=\frac{1}{48}\left(n-n^{-1}\right)+\frac{1}{16n}\,,
\end{equation}
which is the correct dimension in the Ising CFT as $h+\bar{h}$ correctly
reproduces $\frac{1}{2}\frac{1}{12}\left(n-n^{-1}\right)+\frac{1}{8n}$.

We have also seen that, winding the complex fermion field $\chi_{k}(z)=\psi_{k}(z)$
around the branch-point twist field, a phase $e^{i\pi k/n}$ is accumulated
for $k\neq\frac{n}{2}$, which can be attributed to the action of
a $U(1)$ composite twist field. A plausible assumption is that the
decomposition of branch-point twist fields can be rephrased as 
\begin{equation}
\mathcal{T}_{n}^{D}(w,\bar{w})=\mathcal{T}_{\frac{n}{2},n}^{D}(w,\bar{w})\prod_{k=\frac{1}{2}}^{\frac{n-2}{2}}\mathcal{T}_{k,n}^{D}(w,\bar{w})=\mu(w,\bar{w})\prod_{k=\frac{1}{2}}^{\frac{n-2}{2}}\mathcal{O}_{\frac{k}{n}}(w,\bar{w})=\mu(w,\bar{w})\prod_{l=1}^{\frac{n-1}{2}}\mathcal{O}_{\frac{2l-1}{2n}}(w,\bar{w})\,.
\end{equation}
Assuming that this type of factorisation of the $\mathbb{Z}_{2}$
branch-point twist field also holds in the off-critical theory we
can obtain its vacuum expectation value exploiting the results in
Ref. \cite{VEV1} 
\begin{equation}
\langle\mathcal{O}_{\alpha}\rangle=\left(\frac{m}{2}\right)^{\alpha^{2}}\frac{1}{G(1-\alpha)G(1+\alpha)}\,,
\end{equation}
where $G(x)$ is the Barnes G-function. Hence, for the $n$-copy Ising
theory we have 
\begin{equation}
\langle\mathcal{T}_{n}^{D}\rangle=\left(\frac{m}{2}\right)^{\left(\frac{n-n^{-1}}{24}+\frac{1}{8n}-\frac{1}{8}\right)}\langle\mu_{\text{Ising}}\rangle\prod_{l=1}^{\frac{n-1}{2}}\frac{1}{G(1-\frac{2l-1}{2n})G(1+\frac{2l-1}{2n})}\,.
\end{equation}
Using the exact result for $\langle\mu_{\text{Ising}}\rangle$ \cite{MuVEV},
we can write it as 
\begin{equation}
\langle\mu_{\text{Ising}}\rangle=m^{\frac{1}{8}}2^{\frac{1}{12}}e^{-\frac{1}{8}}\mathcal{A}^{\frac{3}{2}}=2^{\frac{1}{4}}\left(\frac{m}{2}\right)^{\frac{1}{8}}\sqrt{\frac{1}{G(\frac{1}{2})G(\frac{3}{2})}\,,}
\end{equation}
and finally we have 
\begin{equation}
\langle\mathcal{T}_{n}^{D}\rangle=2^{\frac{1}{4}}\left(\frac{m}{2}\right)^{\left(\frac{n-n^{-1}}{24}+\frac{1}{8n}\right)}\sqrt{\prod_{l=-\frac{n-1}{2}}^{\frac{n+1}{2}}\frac{1}{G(1-\frac{2l-1}{2n})G(1+\frac{2l-1}{2n})}}\,,
\end{equation}
or, equivalently, using the integral representation 
\begin{equation}
\langle\mathcal{T}_{n}^{D}\rangle=2^{\frac{1}{4}}\left(\frac{m}{2}\right)^{\left(\frac{n-n^{-1}}{24}+\frac{1}{8n}\right)}\exp\left[\int_{0}^{\infty}\frac{\mathrm{d}t}{t}\left(\frac{\sinh t\,\coth\left(\frac{t}{n}\right)-n}{4\sinh^{2}t}-\left(\frac{n-n^{-1}}{24}+\frac{1}{8n}\right)e^{-2t}\right)\right]\,.\label{vev1}
\end{equation}
For $n=1$, this formula equals the vacuum expectation value of the
disorder operator, as obvious. For the less trivial derivative in
$n=1$, we have 
\begin{multline}
\frac{\mathrm{d}}{\mathrm{d}n}\left(m^{-2d_{n}^{D}}\langle\mathcal{T}_{n}^{D}\rangle^{2}\right)\Big|_{n=1}=\left\{ \frac{\ln2}{12}A^{3}2^{\frac{1}{6}}e^{-\frac{1}{4}}+2^{\frac{1}{4}}\exp\left[\int_{0}^{\infty}\frac{\mathrm{d}t}{t}\left(\frac{\cosh t-1}{2\sinh^{2}t}-\frac{1}{4}e^{-2t}\right)\right]\times\right.\\
\left.\times\int_{0}^{\infty}\frac{\mathrm{d}t}{t}\left(\frac{t/\sinh t-1}{2\sinh^{2}t}+\frac{1}{12}e^{-2t}\right)\right\} =-0.111738\ldots\,.\label{vev2}
\end{multline}

\section{Conformal dimensions\label{sec:Appendix-B-CFTDimensions}}

In this appendix we show that Eq. (\ref{eq:ComplexFermionDublet4ptFunctionZ2})
holds for $\mathbb{Z}_{2}$ branch-point twist field in the $c=\frac{1}{2}$
CFT. Let us recall what we want to prove here: 
\begin{equation}
\frac{\langle\psi_{-k}(z)\psi_{k}(z')\mathcal{T}_{n}^{D}(w)\tilde{\mathcal{T}}_{n}^{D}(w')\rangle}{\langle\mathcal{T}_{n}^{D}(w)\tilde{\mathcal{T}}_{n}^{D}(w')\rangle}=\frac{1}{z-z'}\left(\frac{\left(z-w\right)\left(z'-w'\right)}{\left(z-w'\right)\left(z'-w\right)}\right)^{\frac{k}{n}}.\label{eq:ComplexFermionDublet4ptFunctionZ2Appendix}
\end{equation}
The way we proceed is very similar to Refs. \cite{mdc-20b,cs-17}.
We apply the conformal transformation 
\begin{equation}
\xi=\left(\frac{z-w}{z-w'}\right)^{\frac{1}{n}},\label{Mapping}
\end{equation}
which maps the $\mathcal{R}_{n}$ Riemann surface with branch-points
$w$ and $w'$ to the complex plane $\xi\in\mathbb{C}$. After this
uniformising mapping, the twist fields in Eq. (\ref{eq:ComplexFermionDublet4ptFunctionZ2Appendix})
do not disappear, but they become the disorder operator of the
Ising CFT. This is a manifestation of the fact that $\mathcal{T}^{D}$
is the fusion of $\mathcal{T}$ and the disorder field $\mu$. To
check the validity of this idea, we first compute the scaling dimension
of $\mathcal{T}^{D}$ along these lines.

Consider therefore the quantity 
\begin{equation}
\frac{\langle T_{j}(z)\mathcal{T}_{n}^{D}(w)\tilde{\mathcal{T}}_{n}^{D}(w')\rangle}{\langle\mathcal{T}_{n}^{D}(w)\tilde{\mathcal{T}}_{n}^{D}(w')\rangle}\,.\label{TTT}
\end{equation}
After the mapping (\ref{Mapping}), we have 
\begin{multline}
\frac{\langle T_{j}(z)\mathcal{T}_{n}^{D}(w)\tilde{\mathcal{T}}_{n}^{D}(w')\rangle}{\langle\mathcal{T}_{n}^{D}(w)\tilde{\mathcal{T}}_{n}^{D}(w')\rangle}=\frac{\left\langle \left[\left(\frac{\text{d}\xi}{\text{d}z}\right)^{2}T_{j}(\xi)+\frac{c}{12}\left\{ \xi,z\right\} \right]\mu(0)\mu(\infty)\right\rangle }{\langle\mu(0)\mu(\infty)\rangle}\\
%=\frac{c}{12}\left\{ \xi,z\right\} +\left(\frac{\text{d}\xi}{\text{d}z}\right)^{2}\frac{\langleT_{j}(\xi)\mu(0)\mu(\infty)\rangle}{\langle\mu(0)\mu(\infty)\rangle}
=\frac{c}{12}\left\{ \xi,z\right\} +\left(\frac{\text{d}\xi}{\text{d}z}\right)^{2}\frac{\langle\mu(0)T_{j}(\xi)\mu(\infty)\rangle}{\langle\mu(0)\mu(\infty)\rangle}\,,\label{TTTCalculation}
\end{multline}
that can be written as 
\begin{equation}
\begin{split}\frac{\langle T_{j}(z)\mathcal{T}_{n}^{D}(w)\tilde{\mathcal{T}}_{n}^{D}(w')\rangle}{\langle\mathcal{T}_{n}^{D}(w)\tilde{\mathcal{T}}_{n}^{D}(w')\rangle}= & \frac{\left(w-w'\right)^{2}}{\left(z-w\right)^{2}\left(z-w'\right)^{2}}\left[c\frac{1-n^{-2}}{24}+\left(\frac{\xi}{n}\right)^{2}\lim_{\alpha\rightarrow0,\beta\rightarrow\infty}\frac{\langle\mu(\alpha)T_{j}(\xi)\mu(\beta)\rangle}{\langle\mu(\alpha)\mu(\beta)\rangle}\right]\\
= & \frac{\left(w-w'\right)^{2}}{\left(z-w\right)^{2}\left(z-w'\right)^{2}}\left[c\frac{1-n^{-2}}{24}+\left(\frac{\xi}{n}\right)^{2}\lim_{\alpha\rightarrow0,\beta\rightarrow\infty}\frac{1}{16}\frac{\left(\alpha-\beta\right)^{2}}{\left(\alpha-\xi\right)^{2}\left(\xi-\beta\right)^{2}}\right]\\
= & \frac{\left(w-w'\right)^{2}}{\left(z-w\right)^{2}\left(z-w'\right)^{2}}\left[c\frac{1-n^{-2}}{24}+\left(\frac{\xi}{n}\right)^{2}\frac{1}{16}\frac{1}{\xi^{2}}\right]\\
= & \frac{\left(w-w'\right)^{2}}{\left(z-w\right)^{2}\left(z-w'\right)^{2}}\left[c\frac{1-n^{-2}}{24}+\frac{1}{16n^{2}}\right]\,,
\end{split}
\label{TTTCalculation-2}
\end{equation}
where we used \cite{YellowBook} 
\begin{equation}
\frac{\langle\psi(z)\psi(z')\sigma(w)\sigma(w')\rangle}{\langle\sigma(w)\sigma(w')\rangle}=\frac{1}{2}\frac{1}{z-z'}\left[\left(\frac{(z-w)(z'-w')}{(z-w')(z'-w)}\right)^{\frac{1}{2}}+\left(\frac{(z-w')(z'-w)}{(z-w)(z'-w')}\right)^{\frac{1}{2}}\right].\label{PsiPsiSigmaSigma}
\end{equation}
From Eq. (\ref{PsiPsiSigmaSigma}), we also have 
\begin{equation}
\frac{\langle\psi(z)\psi(z')\mu(w)\mu(w')\rangle}{\langle\mu(w)\mu(w')\rangle}=\frac{1}{2}\frac{1}{z-z'}\left[\left(\frac{(z-w)(z'-w')}{(z-w')(z'-w)}\right)^{\frac{1}{2}}+\left(\frac{(z-w')(z'-w)}{(z-w)(z'-w')}\right)^{\frac{1}{2}}\right]\,,\label{PsiPsiMuMu}
\end{equation}
from which $\frac{\langle T(z)\mu(w)\mu(w')\rangle}{\langle\mu(w)\mu(w')\rangle}$
can be obtained. Multiplying the final result by $n$ and comparing
with the Ward identity \eqref{eq:TTDTDTilde}, we find that the right
scaling dimension of the holomorphic part of $\mathcal{T}_{n}^{D}$
which is $\frac{n-n^{-1}}{48}+\frac{1}{16n}$.

Now let us calculate the quantity $\frac{\langle\psi_{-k}(z)\psi_{k}(z')\mathcal{T}_{n}^{D}(w)\tilde{\mathcal{T}}_{n}^{D}(w')\rangle}{\langle\mathcal{T}_{n}^{D}(w)\tilde{\mathcal{T}}_{n}^{D}(w')\rangle}$.
Performing the inverse transformation from $\psi_{k}$ to $\psi_{j}$
and introducing the shorthand $\omega=e^{2\pi i/n}$, we can write
\begin{equation}
\frac{\langle\psi_{-k}(z)\psi_{k}(z')\mathcal{T}_{n}^{D}(w)\tilde{\mathcal{T}}_{n}^{D}(w')\rangle}{\langle\mathcal{T}_{n}^{D}(w)\tilde{\mathcal{T}}_{n}^{D}(w')\rangle}=\sum_{j,j'}\omega^{-(j-1)(k+n/2)}\omega^{(j'-1)(k+n/2)}\frac{\langle\psi_{j}(z)\psi_{j'}(z')\mathcal{T}_{n}^{D}(w)\tilde{\mathcal{T}}_{n}^{D}(w')\rangle}{\langle\mathcal{T}_{n}^{D}(w)\tilde{\mathcal{T}}_{n}^{D}(w')\rangle}\,.
\end{equation}
We are now slightly more cautious with the conformal mapping (\ref{Mapping}),
writing \cite{cs-17} 
\begin{equation}
\xi_{j}=\xi\omega^{j}\,,
\end{equation}
which maps the $j$th sheet of the Riemann surface into a wedge of
angle $2\pi/n$ in $\mathbb{C}$. According to this transformation,
we have 
\begin{multline}
\frac{\langle\psi_{-k}(z)\psi_{k}(z')\mathcal{T}_{n}^{D}(w)\tilde{\mathcal{T}}_{n}^{D}(w')\rangle}{\langle\mathcal{T}_{n}^{D}(w)\tilde{\mathcal{T}}_{n}^{D}(w')\rangle}=\\
=\frac{1}{n}\sum_{j,j'}\left[\omega^{-(j-1)(k+n/2)}\omega^{(j'-1)(k+n/2)}\left(\xi'_{j}(z)\xi'_{j'}(z')\right)^{\frac{1}{2}}\frac{\langle\mu(0)\psi_{j}(\xi_{j})\psi_{j'}(\xi_{j'}')\mu(\infty)\rangle}{\langle\mu(0)\mu(\infty)\rangle}\right]=\\
\frac{1}{n}\sum_{j,j'}\left[\omega^{-(j-1)(k+n/2)}\omega^{(j'-1)(k+n/2)}\left(\xi'_{j}(z)\xi'_{j'}(z')\right)^{\frac{1}{2}}\frac{1}{2}\frac{\sqrt{\xi_{j}(z)/\xi_{j'}(z')}+\sqrt{\xi_{j'}(z')/\xi_{j}(z)}}{\xi_{j}(z)-\xi{}_{j'}(z')}\right]\,,
\end{multline}
where we used Eq. (\ref{PsiPsiSigmaSigma}). We can finally expand
in power series and resum as 
\begin{multline}
\frac{\langle\psi_{-k}(z)\psi_{k}(z')\mathcal{T}_{n}^{D}(w)\tilde{\mathcal{T}}_{n}^{D}(w')\rangle}{\langle\mathcal{T}_{n}^{D}(w)\tilde{\mathcal{T}}_{n}^{D}(w')\rangle}=\\
\frac{1}{n}\sum_{j,j'}\sum_{p=0}^{\infty}\left[\omega^{-(j-1)(k+n/2)-pj}\omega^{(j'-1)(k+n/2)+pj'}\frac{1}{2}\left(\frac{\xi'(z)\xi'(z')}{\xi(z)\xi(z')}\right)^{\frac{1}{2}}\left(\frac{\xi(z')}{\xi(z)}\right)^{p}\right.\\
+\left.\omega^{-(j-1)(k+n/2)-j-pj}\omega^{(j'-1)(k+n/2)+j'+pj'}\frac{1}{2}\left(\frac{\xi'(z)\xi'(z')\xi(z')}{\xi^{3}(z)}\right)^{\frac{1}{2}}\left(\frac{\xi(z')}{\xi(z)}\right)^{p}\right]\\
=n\sum_{q=1}^{\infty}\left[\frac{1}{2}\left(\frac{\xi'(z)\xi'(z')}{\xi(z)\xi(z')}\right)^{\frac{1}{2}}\left(\frac{\xi(z')}{\xi(z)}\right)^{nq-k-n/2}\frac{1}{2}\left(\frac{\xi'(z)\xi'(z')\xi(z')}{\xi^{3}(z)}\right)^{\frac{1}{2}}\left(\frac{\xi(z')}{\xi(z)}\right)^{nq-k-n/2-1}\right]\\
=\frac{n}{\xi^{n}(z)-\xi^{n}(z')}\left[\left(\frac{\xi'(z)\xi'(z')}{\xi(z)\xi(z')}\right)^{\frac{1}{2}}\left(\xi(z')\right)^{n/2-k}\left(\xi(z)\right)^{n/2+k}\right]=\frac{1}{z-z'}\left(\frac{\left(z-w\right)\left(z'-w'\right)}{\left(z-w'\right)\left(z'-w\right)}\right)^{\frac{k}{n}}\,,
\end{multline}
providing the desired result.

\section{Analytic continuation for $f^{D}(\vartheta,n)$
\label{sec:Appendix-C-AnalContForfD}}

The analytic continuation of the quantity $f(\vartheta,n)$ (defined in Eq. (\ref{eq:f(theta,n)}) by replacing $F_{2}^{\mathcal{T}^{D}|1,j}$
with $F_{2}^{\mathcal{T}|1,j}$) was carefully analysed in Ref. \cite{ccd-08}.
It was shown that as the analytic continuation $\tilde{f}(\vartheta,n)$ with domain $n\in[1,\infty)$ can be defined from $f(\vartheta,n)$ for $n=2,3,...$. 
Then $\tilde{f}(\vartheta,n)=f(\vartheta,n)$ for integer $n$ such that $n\geq2$, but for $n\rightarrow1$ we have that 
$f(\vartheta,1)=0$ everywhere except in the origin, where it converges to $\frac{1}{2}$. 
Hence the convergence is non-uniform,
which results in a $\delta$-function in the derivative $\displaystyle\lim_{n\rightarrow1}\frac{\partial}{\partial n}\tilde{f}(\vartheta,n)$,
yielding 
\begin{equation}
\lim_{n\rightarrow1}\frac{\partial}{\partial n}\tilde{f}(\vartheta,n)=\pi^{2}\frac{1}{2}\delta(\vartheta)\,.
\end{equation}
The analysis of \cite{ccd-08} is very detailed, but its full repetition for our case to obtain $\tilde{f}^{D}(\vartheta,1)$
and $\displaystyle\lim_{n\rightarrow1}\frac{\partial}{\partial n}\tilde{f}^{D}(\vartheta,n)$
is not necessary.
We only report some essential ideas for the derivation of $\tilde{f}(\vartheta,n)$ and then discuss
some differences to consider for the $\mathbb{Z}_{2}$ twist field.
First, we recall the definition
\begin{equation}
\langle\mathcal{T}_{n}\rangle^{2}f(\vartheta,n)=\sum_{j=0}^{n-1}F_{2}^{\mathcal{T}|11}(-\vartheta+2\pi i(j))\left(F_{2}^{\mathcal{T}|11}(-\vartheta+2\pi i(j))\right)^{*}=\sum_{j=0}^{n-1}s(\vartheta,j).
\end{equation}
%$\tilde{f}(\vartheta,n)$ can be written as \cite{ccd-08}
%
For the analytic continuation,  we replace $j$ by a continuous variable $z$.
In particular, let us consider the contour integral
\begin{equation}
0=\frac{1}{2\pi i}\oint_{\mathcal{C}}\text{d}z\pi\cot(\pi z)s(\vartheta,z)\,,
\label{contint}
\end{equation}
where the contour is a rectangle with vertices $(-\epsilon-iL,n-\epsilon-iL,n-\epsilon+iL,-\epsilon+iL)$.
This contour integral is zero as when $L\rightarrow\infty$, the
contributions of the horizontal lines vanish and in the Ising model
the vertical contributions cancel each other due to the periodicity
of $s(\vartheta,z+n)=S_{\text{Ising}}^{2}s(\vartheta,z)$ and $S_{\text{Ising}}^{2}=1$.
The integrand has poles at $z=1,2,\ldots,n-1$ and also at $z=\frac{1}{2}\pm\frac{\vartheta}{2\pi i}$
and $z=n-\frac{1}{2}\pm\frac{\vartheta}{2\pi i}$. Evaluating the residues, for real $\vartheta$ we end up with 
\begin{equation}
\sum_{j=1}^{n-1}s(\vartheta,j)=-\tanh\frac{\vartheta}{2}\frac{\text{Im}\left(F_{2}^{\mathcal{T}|11}(-2\vartheta+i\pi,n)-F_{2}^{\mathcal{T}|11}(-2\vartheta+i2\pi n-i\pi,n)\right)}{\langle\mathcal{T}_{n}\rangle}\,,
\end{equation}
and hence the analytic continuation is \cite{ccd-08}
\begin{equation}
\tilde{f}(\vartheta,n)=-\tanh\frac{\vartheta}{2}\frac{\text{Im}\left(F_{2}^{\mathcal{T}|11}(-2\vartheta+i\pi,n)-F_{2}^{\mathcal{T}|11}(-2\vartheta+i2\pi n-i\pi,n)\right)}{\langle\mathcal{T}_{n}\rangle}\,.\label{eq:ftilde}
\end{equation}

We can repeat the same steps for the $\mathbb{Z}_{2}$ twist field. We can write $f^{D}$ as 
\begin{equation}
\langle\mathcal{T}_{n}^{D}\rangle^{2}f^{D}(\vartheta.n)=\sum_{j=0}^{n-1}F_{2}^{\mathcal{T}^{D}|11}(-\vartheta+2\pi ij)\left(F_{2}^{\mathcal{T}^{D}|11}(-\vartheta+2\pi ij)\right)^{*}=\sum_{j=0}^{n-1}s^{D}(\vartheta,j)
\end{equation}
and consider the contour integral
\begin{equation}
\frac{1}{2\pi i}\oint_{\mathcal{C}}\text{d}z\pi\cot(\pi z)s^{D}(\vartheta,z)=-\frac1n\,,
\label{cint2}
\end{equation}
with the same contour as in Eq. \eqref{contint}. Unlike Eq. \eqref{contint}, this integral is non-zero. 
While the vertical contributions again cancel each other, the
horizontal contributions are non zero, because
\begin{equation}
\lim_{L\rightarrow\infty}s^{D}(\vartheta,x\pm iL)=-\frac{1}{n^{2}}, 
\end{equation}
and hence the result is $-\frac{1}{n}$.
We can evaluate the lhs of Eq. \eqref{cint2} by the residue theorem; the poles are at the the same positions as in Eq. \eqref{contint}, i.e. 
$z=1,2,\ldots,n-1$, at $z=\frac{1}{2}\pm\frac{\vartheta}{2\pi i}$, and $z=n-\frac{1}{2}\pm\frac{\vartheta}{2\pi i}$, because the pole
structure of the FFs $F_{2}^{\mathcal{T}^{D}|11}$ and $F_{2}^{\mathcal{T}|11}$ is the same. Evaluating the residues, we end up with
\begin{equation}
\sum_{j=1}^{n-1}s^{D}(\vartheta,j)=-\tanh\frac{\vartheta}{2}\frac{\text{Im}\left(F_{2}^{\mathcal{T}^{D}|11}(-2\vartheta+i\pi,n)+F_{2}^{\mathcal{T}^{D}|11}(-2\vartheta+i2\pi n-i\pi,n)\right)}{\langle\mathcal{T}_{n}^{D}\rangle}-\frac{1}{n}\,,
\end{equation}
from which the analytic continuation is inferred 
\begin{equation}
\tilde{f}^{D}(\vartheta.n)=-\tanh\frac{\vartheta}{2}\frac{\text{Im}\left(F_{2}^{\mathcal{T}^{D}|11}(-2\vartheta+i\pi,n)+F_{2}^{\mathcal{T}^{D}|11}(-2\vartheta+i2\pi n-i\pi,n)\right)}{\langle\mathcal{T}_{n}^{D}\rangle}-\frac{1}{n}\,.\label{eq:fDtilde}
\end{equation}
It is easy to check that $\tilde{f}^{D}(\vartheta,n)=f^{D}(\vartheta,n)$
for odd and integer $n\geq 3$.

The derivative  of $\tilde{f}^{D}(\vartheta,n)$ can be obtained without further work exploiting the property that
the function $\tilde{f}^{D}(\vartheta,n)+\tilde{f}(\vartheta,n)$ is smooth and converges to a smooth function as $n\rightarrow1$.
Indeed, using Eqs. \eqref{eq:ftilde} and \eqref{eq:fDtilde} we immediately have
\begin{equation}
\tilde{f}^{D}(\vartheta,n)+\tilde{f}(\vartheta,n)=\tanh\left(\frac{\theta}{2}\right)\frac{\left(\coth\left(\frac{\theta}{2n}\right)\left(-2\cosh\left(\frac{\theta}{n}\right)+\cos\left(\frac{\pi}{n}\right)+1\right)\right)}{n\left(\cos\left(\frac{\pi}{n}\right)-\cosh\left(\frac{\theta}{n}\right)\right)}-\frac{1}{n}\,,
\end{equation}
and consequently 
\begin{equation}
\begin{split}\lim_{n\rightarrow1}\tilde{f}^{D}(\vartheta,n)+\tilde{f}(\vartheta,n) & =\tanh^{2}\frac{\vartheta}{2},\\
\lim_{n\rightarrow1}\frac{\partial}{\partial n}[\tilde{f}^{D}(\vartheta,n)+\tilde{f}(\vartheta,n)]= & \frac{1}{2}\frac{1-\cosh\vartheta+\frac{2\vartheta}{\sinh\vartheta}}{\cosh^{2}\frac{\vartheta}{2}}\,,
\end{split}
\end{equation}
leading to the main results of this appendix
\begin{equation}
\begin{split}\lim_{n\rightarrow1}\tilde{f}^{D}(\vartheta,n) & =\begin{cases}
\tanh^{2}\frac{\vartheta}{2} & \vartheta\neq0\\
-\frac{1}{2} & \vartheta=0
\end{cases}, \\
\lim_{n\rightarrow1}\frac{\partial}{\partial n}\tilde{f}^{D}(\vartheta,n)= & \frac{1}{2}\frac{1-\cosh\vartheta+\frac{2\vartheta}{\sinh\vartheta}}{\cosh^{2}\frac{\vartheta}{2}}-\pi^{2}\frac{1}{2}\delta(\vartheta).
\end{split}
\label{Ancon}
\end{equation}

We conclude this appendix mentioning the behaviour for $n\rightarrow\infty$, for which we are going to show that 
the limiting functions for $\tilde{f}^{D}(\vartheta,n)$ and $\tilde{f}(\vartheta,n)$ are the same. 
More precisely, we have that
\begin{equation}
\lim_{n\rightarrow\infty}\tilde{f}^{D}(\vartheta,e^{i\phi}n+c)=\frac{\left(2\vartheta^{2}+\pi^{2}\right)\tanh\left(\frac{\vartheta}{2}\right)}{\vartheta\left(\vartheta^{2}+\pi^{2}\right)}\,,
\end{equation}
for any constant $c$ and any direction $\phi$ on the complex plane. 
This large $n$ behaviour is related to the unicity of the analytic continuation \cite{ccd-08} by Carlson's theorem \cite{Carlson}. 
Indeed, let us suppose the existence of another function $\tilde{g}^{D}(\vartheta,n)$, which satisfies $\tilde{g}^{D}(\vartheta,n)=f^{D}(\vartheta,n)$ for odd $n$-s with $n\geq3$. 
We assume that $|\tilde{g}^{D}(\vartheta,n)|<Ce^{q|n|}$ for $\text{Re}(n)>0$ and with $q<\frac{\pi}{2}$; this assumption is motivated
by the fact that both $\text{Tr}\left(\rho_{A}^{n}\right)$ and  $\text{Tr}\left(\rho_{A}^{n}(-1)^{n\hat{Q}_{A}}\right)$ behave so for finite systems, 
see again Ref. \cite{ccd-08} for a detailed discussion.
Then Carlson's theorem can be applied to $\tilde{f}^{D}(\vartheta,n)-\tilde{g}^{D}(\vartheta,n)$
and implies that the difference is identically zero, i.e. the continuation is unique. 
To be more precise, we use Carlson theorem in its standard form \cite{Carlson} by applying it to $\tilde{f}^{D}(\vartheta,2n+1)-\tilde{g}^{D}(\vartheta,2n+1)$,
%because the vanishing of the difference for only odd non-negative
%integers is not sufficient to use the theorem. By the replacement
%$n\rightarrow2n+1$, the vanishing occurs for 
with $n=1, 2,3,4,...$.
The only price to pay is that the growth on the imaginary axis must be bounded by $Ce^{\frac{\pi}{2}|n|}$ rather than 
the usual restriction $Ce^{\pi|n|}$. Anyhow, this is compatible with both the limiting behaviour
of $f^{D}(\vartheta,n)$ and our motivating assumptions for $\tilde{g}^{D}(\vartheta,n)$.

\end{document}